# Acoustic wave polarization and energy flow in periodic beam lattice materials

Andrea Bacigalupo[a,*], Marco Lepidi[b,]

[a]*IMT School for Advanced Studies Lucca, Piazza S. Francesco 19, 55100 Lucca (Italy)*
[b]*DICCA - Dipartimento di Ingegneria Civile, Chimica e Ambientale, Università di Genova, Via Montallegro 1, 16145 Genova (Italy)*

## Abstract

The free propagation of acoustic plane waves through cellular periodic materials is generally accompanied by a flow of mechanical energy across the adjacent cells. The paper focuses on the energy transport related to dispersive waves propagating through non-dissipative microstructured materials. The generic microstructure of the periodic cell is described by a beam lattice model, suitably reduced to the minimal space of dynamic degrees-of-freedom. The linear eigenproblem governing the wave propagation is stated and the complete eigensolution is considered to study both the real-valued dispersion functions and the complex-valued waveforms of the propagating elastic waves. First, a complete family of nondimensional quantities (*polarization factors*) is proposed to quantify the linear polarization or quasi-polarization, according to a proper energetic criterion. Second, a vector variable related to the periodic cell is introduced to assess the directional flux of mechanical energy, in analogy to the Umov-Poynting vector related to the material point in solid mechanics. The physical-mathematical relation between the energy flux and the velocity of the energy transport is recognized. The formal equivalence between the energy and the group velocity is pointed out, according to the mechanical assumptions. Finally, all the theoretical developments are successfully applied to the prototypical beam lattice material characterized by a periodic tetrachiral microstructure. As case study, the tetrachiral material offers interesting examples of perfect and nearly-perfect linear polarization. Furthermore, the nonlinear dependence of the energy fluxes on the elastic waveforms is discussed with respect to the acoustic and optical surfaces featuring the energy spectrum of the material. As final remark, the occurrence of negative refraction phenomena is found to characterize the high-frequency optical surface of the frequency spectrum.

## 1. Introduction

The transport of mechanical energy is a challenging corollary issue related to the free propagation of acoustic waves in microstructured periodic materials. Indeed, the microstructure of the periodic cell can determine significant anisotropies in the material behaviour. As immediate geometric consequence, the dispersion properties of microstructured materials seldom admit pure longitudinal and transversal waveforms, which are typical of isotropic media. More often, the geometric polarization of the acoustic waves is a non-trivial frequency-dependent function of the propagation direction. Consequently, the wavefronts tend to be not-spherical and the direction of the energy fluxes and velocities does not necessarily coincide with the wavevector **k** of the propagating waves [1].

The balance laws governing the transport of mechanical energy are traditional fields of theoretical and applied research, dating back to the pioneer doctoral studies by Nikolay A. Umov in the late nineteenth century [2]. The mechanical energy transferred by acoustic waves, in particular, attracted the attention of eminent scientists all throughout the twentieth century. Among the others, Leon Brillouin studied the mechanical energy transfered between two adjacent cells of a periodic crystal lattice, and related its flux density to the particle velocities through the concept of characteristic impedance [3, 4]. Almost concurrently, Maurice A. Biot established some general theorems relating the group wave velocity and the energy velocity in anisotropic, non-homogeneous, non-dissipative media [5]. Over the last decades, specific issues related to the transport of mechanical energy have been treated in different monographs concerning anisotropic elastic solids [6], viscoelastic heterogeneous solids and fluids [7], anisotropic, anelastic, porous and electromagnetic materials [8], viscoelastic layered media [9]. Occasional but sharp attention has been specifically devoted to the flow of mechanical energy in periodic systems, including crystal lattices [10, 11] and structural assemblies [12, 13].

The essential idea, shared by the largest majority of literature studies, is that strict formal and substantial analogies can be established between the radiation of electromagnetic energy and the transfer of mechanical energy [14]. According to this standpoint, the well-known *Poynting vector* associated to the electromagnetic field can systematically be replaced by the *Umov-Poynting vector* **s** in solid mechanics [2, 15]. In a continuous material the Umov-Poynting vector is a local quantity, expressing the mechanical power density related to the velocity field of a certain natural motion. In the specific case of natural wave motions, the Umov-Poynting vector depends on the dispersion relation $\omega(\mathbf{k})$ and is a quadratic function of the waveform $\boldsymbol{\psi}(\mathbf{k})$. Most significantly, the **s**-vector accounts also for the directional density (per unit surface with outward unit normal **n**) of the mechanical power flux $\mathbf{s}\cdot\mathbf{n}$ flowing through a certain material point undergoing harmonic wave oscillations. As primary consequence, the Umov-Poynting vector can directly be related to the direction and velocity of the mechanical energy transferred by the acoustic waves traveling through the medium.

[*]Corresponding author
 *Email address:* andrea.bacigalupo@imtlucca.it (Andrea Bacigalupo)

Based on the above considerations, the Umov-Poynting vector is a valuable mathematical tool in a variety of research fields, ranging from the science of energy focusing and harvesting to the engineering of smart materials and metamaterials. Considering – in particular – the dispersion properties of materials, many literature studies have investigated how different constitutive properties (namely anisotropy, viscoelasticity, non-linearity, non-locality) modify the relation between the energy velocity and the group velocity of harmonic waves [1, 16–19]. In this respect, it may be interesting to recall an early intuition by Ivan Tolstoy [20] (and emphasized by Biot [5]), about the computational advantages of an energy-based assessment of the group velocity. Indeed, the group velocity can be assessed by inverting its analytical relation with the velocity of the energy flux. This indirect procedure allows to by-pass the computational bottleneck related to numerically evaluating the dispersion function derivative $\nabla_\mathbf{k}\omega(\mathbf{k})$. For all these motivations, the Umov-Poynting vector is gaining increasing attention in many contemporary works specifically devoted to the dispersion characterization of *periodic* materials [21–24]. Indeed, studying and governing the direction of the power flux is becoming a key aspect in several advanced applications, like the fine tuning of phononic filters [25–29], the elastic wave beaming and funneling [30–32] or the acoustic focusing, lensing and cloaking [33–36]. Among the other possibilities concerned with periodic materials, a fascinating prospect is the functional design of negative refraction properties ($\mathbf{s}\cdot\mathbf{k} < 0$) in phononic crystals [37] and in square chiral lattices [38].

Although the physical concept and the mathematical role of the Umov-Poynting vector are well-established aspects in continuum mechanics, the proper definition of its counterpart for *periodic heterogeneous* materials may still offer some open and challenging tasks. The first, quite natural option is to somehow circumvent the issue by adopting equivalent homogeneous continua [39–47]. By virtue of proper averaging procedures for the microscopic fields, the homogenization techniques filter out the irrelevant local fluctuations of the field variables around the periodic heterogeneities. This powerful methodological approach is quite popular in solid mechanics as well as in electromagnetism. Within the framework of Maxwell's theory, the leading idea is to homogenize the electric and magnetic fields. Roughly, this result is achievable by spatially averaging the field variables over the heterogeneous domain of the periodic cell. Ultimately, the homogenized electromagnetic fields allow a consistent formulation of the macroscopic (spatially-averaged) Poynting vector in periodic heterogeneous continua, such as lossless stratified media [48], lossy structured media [49] and negative-index metamaterials [50]. Within the parallel framework of solid mechanics, the homogenization techniques are adopted to describe the average flux of mechanical elastic energy flowing in periodic micro-architectured materials. The macroscopic strain and stress fields of a suited energetically-equivalent continuum can be employed to the purpose. Following this approach, macroscopic quantities analogous to the Umov-Poynting vector can be introduced in homogenized classic or non-classic continua, such a strain-gradient anisotropic media [51]. It is worth remarking that, after the homogenization, the macroscopic equivalent of the Umov-Poynting vector defines a continuous vector field, that is, a directional density of mechanical power can be attributed to each material point of the homogenized solid.

As a valid alternative to homogenization, the flux of mechanical energy in heterogeneous materials can be evaluated according to lagrangian dynamic formulations. Specific issues related to the energy transport in periodic microstructured materials can be addressed by adopting crystal lattice models [10] or beam lattice models [13, 52]. According to this methodological approach, each periodic cell is regarded as a multi-atomic unit (molecule), whose configuration is described by a finite number of lagrangian coordinates. According to the basic crystal lattice model, purely attractive and repulsive potentials are defined to determine the reciprocal interactions between pairs of point masses (atoms). Long-term forces can be considered by establishing intermolecular interactions. According to the simplest beam lattice model, instead, axial-bending elastic potentials are defined to describe the elastic coupling between close pairs of orientable massive points (nodes). It is worth remarking that, within the lattice framework, the Umov-Poynting vector defines a discrete vector field, that is, the directional density of mechanical power must be attributed to a minimal reference unit, naturally coincident with the periodic cell.

The objective of the present work is to outline a general theoretical framework to investigate the transport of mechanical energy related to the propagation of dispersive acoustic waves in certain periodic microstructured materials. Specifically, the paper focuses on lagrangian formulations of non-dissipative beam lattice models, whose finite dimension can conveniently be reduced by classic procedures of quasi-static condensation (Section 2). The linear eigenproblem governing the free propagation of harmonic plane waves is stated for the generic microstructure (Paragraph 2.1). Specifically, energy-based mathematical tools are defined to quantify the linear polarization of the waveforms (Paragraph 2.2). Therefore, considering a dispersive wave propagating with generic polarization, a vector variable describing the directional flux of the transferred mechanical energy is defined (Paragraph 2.3). A mathematical relation between the velocity of the energy flux and the spectral properties of the condensed beam lattice model is established (Paragraph 2.4). Finally, the energy velocity is demonstrated coincident with the group velocity, which is assessed by means of different alternative formulations (Paragraph 2.5). All the theoretical developments are successfully applied to the tetrachiral periodic material (Section 3). For this specific case-study, perfect and quasi-perfect polarization of the waveforms are quantitatively evaluated, and negative refractions are recognized for the high-frequency optical surface of the dispersion spectrum. Concluding remarks are finally pointed out.

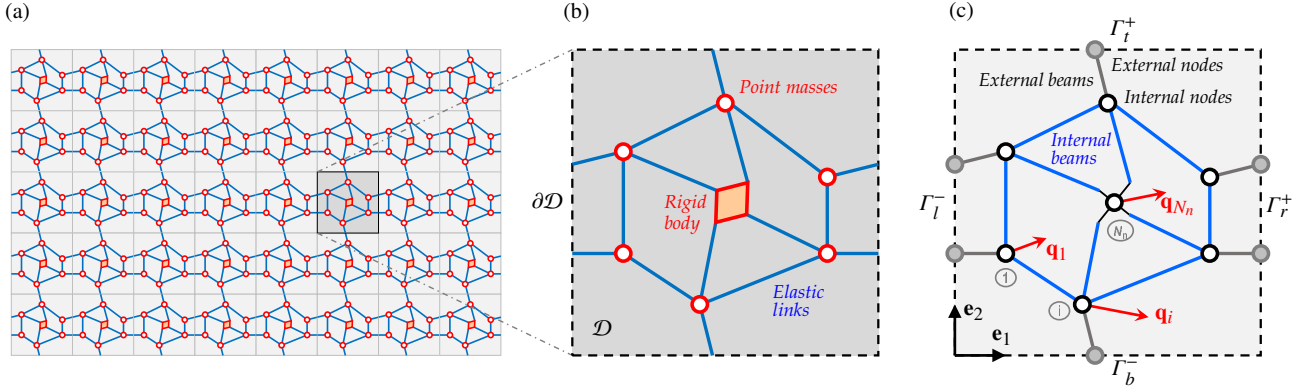

Figure 1: Two-dimensional periodic material (a) repetitive planar pattern, (b) periodic cell, (c) beam lattice model.

## 2. Beam lattice model

The mechanical behaviour of many periodic microstructured solids can be described – in its essence – by a geometrically regular pattern of material points (or bodies), coupled by linear elastic links with extensional and/or flexural stiffness. Therefore, a *beam lattice* model can be formulated by employing the material points as microstructural rigid nodes, and modeling the elastic inter-node connections as flexible massless beams.

Depending on the microstructural periodicity, the geometric domain $\mathcal{D}$ of the elementary cell may host one or more *internal* nodes, connected by *internal* beams. By virtue of the cellular symmetries, the closed boundary $\partial\mathcal{D}$ of the $\mathcal{D}$-domain usually crosses the midspan of all the *external* beams, providing the elastic connection between two adjacent cells. Therefore, the *microstructural boundary* $\Gamma$ of the elementary cell is composed by a certain number of massless *external* nodes, lying at the intersection between the $\partial\mathcal{D}$-boundary and the external beams.

According to the beam lattice formulation, the dynamic response of the periodic cell is governed by a multi-degrees-of-freedom lagrangian model, referred to a finite number $N_n$ of internal and external nodes. In two-dimensional periodic materials, the in-plane configuration of the $i$-th node can be described by three nondimensional components of motions, namely the displacements $u_i, v_i$ (along the coordinate in-plane directions $\mathbf{e}_1, \mathbf{e}_2$) and the rotation $\theta_i$ (along the out-of-plane direction $\mathbf{e}_3$ according to the right-hand rule). Therefore, the actual cell configuration is completely described by a generalized displacement vector $\mathbf{q} = (\mathbf{q}_1, ..., \mathbf{q}_{N_n})$, collecting column-wise all the nodal components of motion $\mathbf{q}_i = (u_i, v_i, \theta_i)$ for $i = 1, .., N_n$.

Within the compass of conservative materials, the free dynamics of the periodic cell is governed by an ordinary differential system of nondimensional equations of motion

$$\mathbf{M}(\mathbf{p})\ddot{\mathbf{q}} + \mathbf{K}(\mathbf{p})\mathbf{q} = \mathbf{f} \qquad (1)$$

where $\mathbf{M}(\mathbf{p})$ and $\mathbf{K}(\mathbf{p})$ are the symmetric mass and stiffness matrices, depending on a minimal set $\mathbf{p}$ of independent mechanical parameters. The dot indicates differentiation with respect to the nondimensional time $\tau = \Omega_c t$, where $\Omega_c$ stands for a suited known dimensional frequency. Finally, the vector $\mathbf{f}$ collects the elastic forces exerted by the adjacent cells to the external nodes.

The $N$-dimensional configuration vector ($N = 3N_n$) can conveniently be partitioned $\mathbf{q} = (\mathbf{q}_a, \mathbf{q}_p)$ by distinguishing

- the *active* displacements $\mathbf{q}_a$ of the massive internal nodes, where inertial forces may develop
- the *passive* displacements $\mathbf{q}_p$ of the massless external nodes, where only static forces may act.

with $N_a$ and $N_p$ (where $N_a + N_p = N$) indicating the dimension of the displacement vectors $\mathbf{q}_a$ and $\mathbf{q}_p$, respectively.

According to the displacement partition (and dropping the matrix dependence on $\mathbf{p}$), the equations of motion read

$$\begin{bmatrix} \mathbf{M}_a & \mathbf{O} \\ \mathbf{O} & \mathbf{O} \end{bmatrix} \begin{pmatrix} \ddot{\mathbf{q}}_a \\ \ddot{\mathbf{q}}_p \end{pmatrix} + \begin{bmatrix} \mathbf{K}_{aa} & \mathbf{K}_{ap} \\ \mathbf{K}_{pa} & \mathbf{K}_{pp} \end{bmatrix} \begin{pmatrix} \mathbf{q}_a \\ \mathbf{q}_p \end{pmatrix} = \begin{pmatrix} \mathbf{0} \\ \mathbf{f}_p \end{pmatrix} \qquad (2)$$

where $\mathbf{O}$ stands for empty matrices and the rectangular matrices $\mathbf{K}_{ap} = \mathbf{K}_{pa}^\top$ account for the elastic coupling between the internal and external nodes. Due to the absence of inertial forces, the lower equation expresses the quasi-static equilibrium regulating the passive displacements of the external nodes. For the sake of generality, it is worth remarking that the presence of active nodes located at the microstructural boundary (for instance due to a different choice of the periodic cell) can be treated with an analogous procedure, without conceptual difficulties.

### 2.1. Wave propagation

The acoustic wave propagation through the two-dimensional material can be analyzed by exploiting the periodicity of the cell microstructure, according to the Floquet-Bloch theory. If a rectangular $\mathcal{D}$-domain is considered, the microstructural boundary $\Gamma$ can conventionally be segmented into a *negative* and a *positive* sub-boundaries denoted $\Gamma^-$ and $\Gamma^+$. The negative sub-boundary $\Gamma^-$ includes all the external nodes belonging to the left $\Gamma_l^-$ and bottom $\Gamma_b^-$ segments of the rectangular boundary. The positive sub-boundary $\Gamma^+$ includes instead all the external nodes belonging to the right $\Gamma_r^+$ and top $\Gamma_t^+$ segments of the rectangular boundary (Figure 1c).

According to the segmentation of the $\Gamma$-boundary, the vector collecting the passive displacements can be partitioned $\mathbf{q}_p = (\mathbf{q}_p^-, \mathbf{q}_p^+)$, where $N_p^-$ and $N_p^+$ indicate the dimensions of the subvectors $\mathbf{q}_p^-$ and $\mathbf{q}_p^+$, respectively (with $N_p^- + N_p^+ = N_p$ and $N_p^- = N_p^+$, for the periodicity). Furthermore, the displacement subvectors $\mathbf{q}_p^-$ and $\mathbf{q}_p^+$ can properly be partitioned $\mathbf{q}_p^- = (\mathbf{q}_l^-, \mathbf{q}_b^-)$ and $\mathbf{q}_p^+ = (\mathbf{q}_r^+, \mathbf{q}_t^+)$, where $N_l, N_b, N_r, N_t$ indicate the dimensions of the subvectors $\mathbf{q}_l^-, \mathbf{q}_b^-, \mathbf{q}_r^+, \mathbf{q}_t^+$ (with $N_l = N_r$ and $N_b = N_t$). The same partition is applied to the force vector $\mathbf{f}_p$.

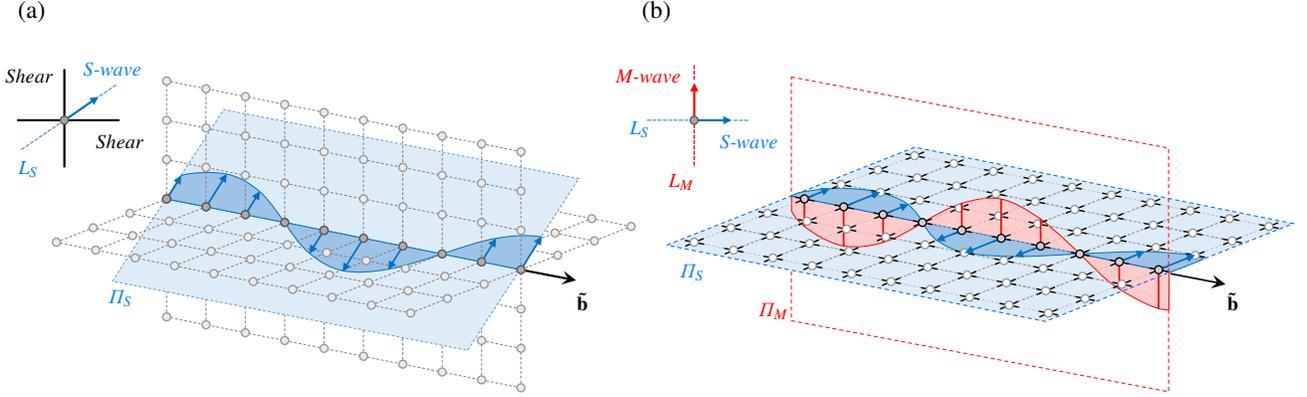

Figure 2: Linear polarization of elastic waves in periodic lattices: (a) three-dimensional lattice of point masses, (b) two-dimensional beam lattice.

Within this framework, the elastic waves propagating across adjacent cells can be described by invoking the Floquet-Bloch theory for periodic systems [53]. To this purpose, the nondimensional wavenumbers $\beta_1$ and $\beta_2$ of the planar waves propagating along the coordinate axes can be defined. Then, the nondimensional wavevector $\mathbf{b} = (\beta_1, \beta_2)$ can be introduced to span the first Brillouin zone $\mathcal{B}$ related to the $\mathcal{D}$-domain. Therefore, the Floquet-Bloch conditions can be imposed on the passive displacements (and forces) in the form $\mathbf{q}_p^+ = \mathbf{L}(\mathbf{b})\mathbf{q}_p^-$ and $\mathbf{f}_p^+ = -\mathbf{L}(\mathbf{b})\mathbf{f}_p^-$, where the $\mathbf{b}$-dependent matrix $\mathbf{L}(\mathbf{b})$ is diagonal and complex-valued (see the AppendixA.1).

Moreover, a convenient reduction of the model dimension can be pursued by applying a classic quasi-static condensation to the *passive* displacements. This procedure univocally relates the passive displacements and forces to the active displacements according to the linear laws $\mathbf{q}_p = \mathbf{S}_{pa}^- \mathbf{q}_a$ and $\mathbf{f}_p^- = \mathbf{F}_{pa}^- \mathbf{q}_a$, without approximation. Finally, the free wave dynamics of the periodic material is governed by the reduced equation

$$\mathbf{M}_a(\mathbf{p})\ddot{\mathbf{q}}_a + \mathbf{K}_a(\mathbf{p}, \mathbf{b})\mathbf{q}_a = \mathbf{0} \quad (3)$$

where the active degrees-of-freedom $\mathbf{q}_a$ play the role of lagrangian coordinates and $\mathbf{K}_a(\mathbf{p}, \mathbf{b})$ is a $\mathbf{b}$-dependent generalization of the condensed stiffness matrix, with hermitian properties. The general form of the stiffness matrix $\mathbf{K}_a(\mathbf{p}, \mathbf{b})$ and the auxiliary matrices $\mathbf{S}_{pa}^-$, $\mathbf{F}_{pa}^-$ are reported in the AppendixA.2.

The wave equation (3) can be tackled by imposing the harmonic mono-frequent solution $\mathbf{q}_a = \boldsymbol{\psi}_a \exp(\iota\omega\tau)$, where $\omega = \Omega/\Omega_c$ and $\Omega$ are the unknown nondimensional and dimensional wave frequency, respectively. It is worth remarking that, in combination with the assigned Floquet-Bloch conditions, this solution describes forward propagating waves. Therefore, eliminating the dependence on time gives the linear eigenproblem

$$\left(\mathbf{K}_a(\mathbf{p}, \mathbf{b}) - \lambda \mathbf{M}_a(\mathbf{p})\right)\boldsymbol{\psi}_a = \mathbf{0} \quad (4)$$

which can be reformulated in the equivalent *standard* form by decomposing the mass matrix $\mathbf{M}(\mathbf{p}) = \mathbf{Q}(\mathbf{p})^\top \mathbf{Q}(\mathbf{p})$, yielding

$$\left(\mathbf{H}(\mathbf{p}, \mathbf{b}) - \lambda \mathbf{I}\right)\boldsymbol{\phi}_a = \mathbf{0} \quad (5)$$

where $\lambda = \omega^2$ and $\mathbf{H}(\mathbf{p}, \mathbf{b}) = \mathbf{Q}(\mathbf{p})^{-\top} \mathbf{K}_a(\mathbf{p}, \mathbf{b}) \mathbf{Q}(\mathbf{p})^{-1}$. The complete eigensolution includes $N_a$ real-valued eigenvalues $\lambda(\mathbf{p}, \mathbf{b})$ and the corresponding complex-valued *standard eigenvectors* $\boldsymbol{\phi}_a(\mathbf{p}, \mathbf{b})$. The eigenvalues $\lambda$ are the zeros of the characteristic function $F(\lambda, \mathbf{p}, \mathbf{b}) = \det(\mathbf{H}(\mathbf{p}, \mathbf{b}) - \lambda \mathbf{I})$. The *active* waveforms $\boldsymbol{\psi}_a(\mathbf{p}, \mathbf{b})$ of the $\omega(\mathbf{p}, \mathbf{b})$-monofrequent propagating wave are related to the standard eigenvectors by the relation $\boldsymbol{\psi}_a = \mathbf{Q}^{-1}\boldsymbol{\phi}_a$. The condensation relations $\boldsymbol{\psi}_p^- = \mathbf{S}_{pa}^- \boldsymbol{\psi}_a$ and $\boldsymbol{\psi}_p^+ = \mathbf{LS}_{pa}^- \boldsymbol{\psi}_a$ express the *passive* waveforms $\boldsymbol{\psi}_p$ as a linear function of the active waveforms $\boldsymbol{\psi}_a$. Similarly, the relations $\boldsymbol{\varphi}_p^- = \mathbf{F}_{pa}^- \boldsymbol{\psi}_a$ and $\boldsymbol{\varphi}_p^+ = -\mathbf{L}\mathbf{F}_{pa}^- \boldsymbol{\psi}_a$ determine the form of the passive forces. The dispersion functions $\omega(\mathbf{b})$ over the first Brillouin zone $\mathcal{B}$ fully characterize the Floquet-Bloch spectrum (or band structure) of the material described by certain mechanical parameters $\mathbf{p}$.

### 2.2. Wave polarization

The polarization state is a geometric property characterizing the electromagnetic radiations. If a monochromatic electromagnetic wave radiates in the three-dimensional free space, the synchronous electric and magnetic fields oscillate transversally to the propagation direction, along two mutually orthogonal directions. Longitudinal oscillations, parallel to the propagation direction, are not allowed. If the transversal oscillation direction of the electric field is time-independent, the wave is said to be *linearly polarized*, and the oscillation direction is referred to as *polarization line*. The plane defined by the polarization line and the propagation direction is known as *polarization plane*.

In analogy with the electromagnetic polarization, the *acoustic polarization* is a dispersion property of the vibration waves propagating through an elastic medium. In a three-dimensional lattice of point masses (Figure 2a), transversal waves (*shear S-waves*) can co-exist with longitudinal waves (*compression P-waves*). Coupled or hybrid waves, whose waveforms combine transversal and longitudinal oscillations, can also occur. Rigorously, only transversal (shear) waves can be *polarized*, that is, geometrically oriented along a certain polarization line $L_S$, orthogonal to the propagation direction $\tilde{\mathbf{b}}$. The polarization line and the propagation direction identify the *polarization plane* $\Pi_S$. Conversely, longitudinal (compression) waves cannot be polarized, because a polarization plane cannot be identified.

In the specific framework of two-dimensional beam lattices (Figure 2b), the wave motion is confined in the material plane. Nonetheless, the linear wave polarization can occur in two principal forms. First, the beam lattice can develop in-plane shear $S$-waves, vibrating orthogonally to the propagation direction (blue wave in Figure 2b). The corresponding polarization line $L_S$ define the polarization plane $\Pi_S$, which coincides with the material plane. Second, the beam lattice is composed of orientable rigid nodes. The oscillations of their rotational degrees of freedom allow the propagation of rotation, or moment $M$-waves (red wave in Figure 2b), fully decoupled by the shear $S$-waves. The corresponding polarization line $L_M$ defines the polarization plane $\Pi_M$, which is orthogonal to the material plane. Finally, non-polarized longitudinal $P$-waves can also occur.

The acoustic waves propagating in a beam lattice are generally not perfectly polarized. Indeed, the generic propagation direction, fixed by the unit vector $\tilde{\mathbf{b}} = \mathbf{b}/\|\mathbf{b}\|$ corresponds to coupled or hybrid waves, whose waveforms are simultaneously participated by transversal, longitudinal and rotational oscillations. Consequently, a first basic task may be to immediately recognize the exceptional occurrence of (perfect) wave polarization. Therefore, if perfect polarization occurs, a second task may be to clearly identify the qualitative nature (shear or moment) of the polarized wave. Finally, if perfect polarization does not occur, a third advanced task may be to quantify the degree of quasi-polarization (or, equivalently, the defect of perfect polarization) of a certain waveform.

With respect to the existing literature on the topic, here the novel idea is to approach this threefold task with a unique mathematical tool, by introducing a pair of *polarization factors*

- Shear factor $\qquad \Lambda_S = \dfrac{(\mathbf{R}_S \mathbf{B} \boldsymbol{\phi}_a)^\dagger (\mathbf{R}_S \mathbf{B} \boldsymbol{\phi}_a)}{\boldsymbol{\phi}_a^\dagger \boldsymbol{\phi}_a}$ (6)

- Moment factor $\qquad \Lambda_M = \dfrac{(\mathbf{R}_M \mathbf{B} \boldsymbol{\phi}_a)^\dagger (\mathbf{R}_M \mathbf{B} \boldsymbol{\phi}_a)}{\boldsymbol{\phi}_a^\dagger \boldsymbol{\phi}_a}$ (7)

where the boolean matrices $\mathbf{R}_S = [\mathrm{diag}(\mathbf{e}_2), ..., \mathrm{diag}(\mathbf{e}_2)]$ and $\mathbf{R}_M = [\mathrm{diag}(\mathbf{e}_3), ..., \mathrm{diag}(\mathbf{e}_3)]$ are diagonal, while the symbol $()^\dagger$ stands for the transpose and complex conjugate. The skew-symmetric matrix $\mathbf{B}$ is a $\tilde{\mathbf{b}}$-dependent rotation matrix, which can be built up to align the $\boldsymbol{\phi}_a$-wavecomponents with the propagation direction, as reported in the AppendixA.

From the mathematical viewpoint, the polarization factors are quadratic functions of the standard waveform $\boldsymbol{\phi}_a$ and can attain all the real values in the range [0, 1]. It is worth remarking that the polarization factors are independent of the particular waveform normalization, by construction. From the mechanical viewpoint, the $\Lambda_S$-value and the $\Lambda_M$-value quantify the degree of wave polarization in the planes $\Pi_S$ and $\Pi_M$, respectively (Figure 3). Perfectly polarized $S$-waves have unitary shear factor ($\Lambda_S = 1$) and null moment factor ($\Lambda_M = 0$). Conversely, perfectly polarized $M$-waves have unitary moment factor ($\Lambda_M = 1$) and null shear factor ($\Lambda_S = 0$). The complementary

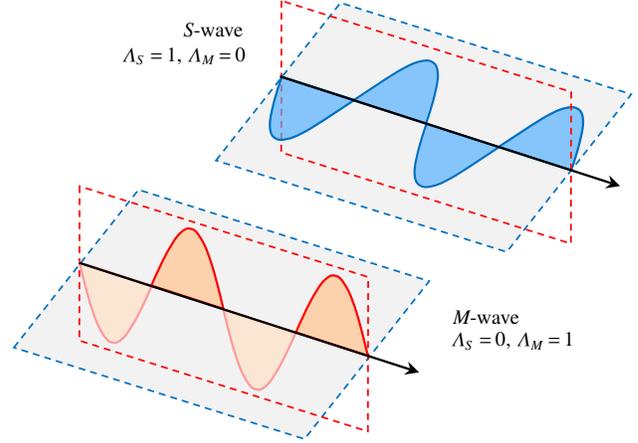

Figure 3: Perfectly polarized shear ad moment waves propagating through a two-dimensional beam lattice material.

*compression factor* $\Lambda_P = 1 - \Lambda_S - \Lambda_M$ can also be defined to identify non-polarized, perfectly longitudinal compression $P$-waves ($\Lambda_P = 1$). Clearly, polarization factors close to unity correspond to *nearly-polarized* $S$-waves ($\Lambda_S \simeq 1$) or $M$-waves ($\Lambda_M \simeq 1$). Conversely, hybrid waves have comparable polarization factors, indicating that two or more components of motion (shear, moment, compression) participate in the beam lattice oscillations in a non-negligible manner.

An immediate physical interpretation can be recognized for the polarization factors, based on the mechanical energy of the propagating wave. Indeed, recalling the relation $\boldsymbol{\psi}_a = \mathbf{Q}^{-1} \boldsymbol{\phi}_a$, the localization factors can be expressed

- Shear factor $\qquad \Lambda_S = \dfrac{\boldsymbol{\psi}_a^\dagger \mathbf{R}_S^\top \mathbf{M}_a \mathbf{R}_S \boldsymbol{\psi}_a}{\boldsymbol{\psi}_a^\dagger \mathbf{M}_a \boldsymbol{\psi}_a}$ (8)

- Moment factor $\qquad \Lambda_M = \dfrac{\boldsymbol{\psi}_a^\dagger \mathbf{R}_M^\top \mathbf{M}_a \mathbf{R}_M \boldsymbol{\psi}_a}{\boldsymbol{\psi}_a^\dagger \mathbf{M}_a \boldsymbol{\psi}_a}$ (9)

where the relation $\mathbf{B}^\top \mathbf{B} = \mathbf{I}$ has been used. Therefore, the polarization factors $\Lambda_S$ and $\Lambda_M$ can be recognized as the ratio of the kinetic energy stored in one or the other polarization plane ($\Pi_S$ and $\Pi_M$) with respect to the total kinetic energy of a harmonically propagating wave. Similarly, the compression factor $\Lambda_P = 1 - \Lambda_S - \Lambda_M$ expresses the complementary fraction of total kinetic energy, which remains in the propagation direction. In this respect, the polarization factors have close similarity with the *localization factors* currently employed in the classic modal analysis of periodic or quasi-periodic structures [54, 55].

It is worth noting that, if the propagation direction $\tilde{\mathbf{b}}$ coincides with one of the coordinate axes identified by the unit vectors $\mathbf{e}_1$, the localization factors assume the simplified form

$$\Lambda_S = \dfrac{(\mathbf{R}_S \boldsymbol{\phi}_a)^\dagger (\mathbf{R}_S \boldsymbol{\phi}_a)}{\boldsymbol{\phi}_a^\dagger \boldsymbol{\phi}_a}, \qquad \Lambda_M = \dfrac{(\mathbf{R}_M \boldsymbol{\phi}_a)^\dagger (\mathbf{R}_M \boldsymbol{\phi}_a)}{\boldsymbol{\phi}_a^\dagger \boldsymbol{\phi}_a} \qquad (10)$$

and similar simplifications can be derived for wave propagating along the other coordinate axes identified by the unit vectors $\mathbf{e}_2$.

## 2.3. Energy flux

In electrodynamics, the Poynting Theorem [15] states a variant of the energy conservation law. According to the integral form of this theorem, the decrease rate of the electromagnetic energy $\mathcal{E}$ stored in a finite control volume $\mathcal{V}$, bounded by the closed surface $\mathcal{S}$ with unit outward normal $\mathbf{n}$, is

$$-\dot{\mathcal{E}} = \int_{\mathcal{S}} (\mathfrak{s} \cdot \mathbf{n})\, dS + \int_{\mathcal{V}} (\mathfrak{E} \cdot \mathfrak{J})\, dV \quad (11)$$

where $\mathfrak{E}$ and $\mathfrak{J}$ stand for the electric field and the current density, respectively, while $\mathfrak{s}$ is the *Poynting vector*

$$\mathfrak{s} = \frac{1}{\mu} \mathfrak{E} \times \mathfrak{B} \quad (12)$$

where $\mathfrak{B}$ is the magnetic flux density, $\mu$ is the magnetic permittivity and $\times$ stands for the vector product. The physical interpretation is that the energy decrease per unit time is equal to the energy lost in the control region of space due to the Joule effect ($\mathcal{V}$-volume integral), plus the energy flux leaving that region through its boundary ($\mathcal{S}$-surface integral). Consequently, the Poynting vector can be interpreted as an energy flux density or – rigorously – a directional energy flux per unit surface.

In adiabatic conditions, a non-dissipative solid obeys to the conservation principle of the mechanical energy $\mathcal{E}$. In analogy with the Poynting theorem in the electrodynamic theory, the Umov-Poynting Theorem [2] is a variant of the balance law for the mechanical energy. In the standard form, this law states that the variation rate of mechanical energy $\dot{\mathcal{E}}$ is equal to the external mechanical power $\mathcal{P}_E$ [7, 56]. In the integral form, the Umov-Poynting Theorem expresses the decrease rate of mechanical energy in a finite material volume $\mathcal{V}$, bounded by the closed surface $\mathcal{S}$ with unit outward normal $\mathbf{n}$, as

$$-\dot{\mathcal{E}} = \int_{\mathcal{S}} (\mathbf{s} \cdot \mathbf{n})\, dS + \int_{\mathcal{V}} ((\nabla \cdot \mathbf{T} - \varrho \ddot{\mathbf{u}}) \cdot \dot{\mathbf{u}})\, dV \quad (13)$$

where $\mathbf{u}$ is the displacement field, $\mathbf{T}$ is the stress tensor, $\varrho$ is the mass density, and finally $\mathbf{s}$ is the *Umov-Poynting vector*

$$\mathbf{s} = -\mathbf{T}\dot{\mathbf{u}} \quad (14)$$

which represents the directional flux of mechanical energy per unit $\mathcal{S}$-surface. The equation (13) assumes an immediate physical interpretation in the *free oscillations* regime ($\mathcal{P}_E = 0$): the mechanical energy in the material volume is conserved in time ($\dot{\mathcal{E}} = 0$) if, first, the increment rate of the internal energy (with density $(\nabla \cdot \mathbf{T}) \cdot \dot{\mathbf{u}}$) is fully balanced by the decrement rate of the kinetic energy (with density $(\varrho \ddot{\mathbf{u}}) \cdot \dot{\mathbf{u}}$) and, second, the total energy transfer across the boundary (the $\mathcal{S}$-surface integral) is null. The derivation of the equation (13) from the balance law of the mechanical energy is reported in the AppendixB.1.

The mathematical definition of the Umov-Poynting vector for homogeneous continua can properly be adapted to study the mechanical energy flux through periodic solids and beam lattices. Considering first a generic periodic solid, the elementary cell domain $\mathcal{D}$ can be assumed as reference material volume.

Following the Umov-Poynting Theorem, the total flux of mechanical energy flowing out of the cell boundary $\partial\mathcal{D}$ is

$$J = \int_{\partial\mathcal{D}} (\mathbf{s} \cdot \mathbf{n})\, dS \quad (15)$$

where, in the absence of internal body forces and considering that the normal $\mathbf{n}$ is necessarily determined (by the geometric assumption of the cell domain as reference volume), the integral kernel can be conveniently expressed by the important relation

$$\mathbf{s} \cdot \mathbf{n} = -\mathbf{f} \cdot \dot{\mathbf{u}} \quad (16)$$

which relates the Umov-Poynting vector $\mathbf{s}$ to the velocity field $\dot{\mathbf{u}}$ of the cell boundary and the forces $\mathbf{f}$ exerted by the adjacent cells (see also the AppendixB.1). Consequently, for a given natural motion $\mathbf{u}(t)$, the flux density of the mechanical energy $\mathbf{s} \cdot \mathbf{n}$ is a local dynamic quantity, defined all along the closed cellular boundary. Locally, it is equal to the (opposite of the) dot product between the force $\mathbf{f}$ acting on a certain moving point of the $\partial\mathcal{D}$-boundary, and the point velocity $\dot{\mathbf{u}}$. It is worth noting that, in the *free oscillations* regime, the Umov-Poynting vector establishes a solenoidal field (see also the AppendixB.1). Consequently, the flux of mechanical energy flowing out of any closed boundary, including the $J$-flux flowing out of the cellular boundary $\partial\mathcal{D}$ in the equation (15), is identically null.

Considering a beam lattice, the closed boundary $\partial\mathcal{D}$ must be replaced by the entire microstructural $\Gamma$-boundary of the elementary cell. Similarly, the equations (15)-(16) for the mechanical energy flux can be adapted by replacing the integral formulation with the dot product between the passive forces $\mathbf{f}_p$ and the passive velocities $\dot{\mathbf{q}}_p$ of all and only the external massless nodes of the cell microstructure. Consequently, the total flux of mechanical energy flowing out of the microstructural $\Gamma$-boundary of the periodic cell can be determined as

$$J = -\mathbf{f}_p \cdot \dot{\mathbf{q}}_p = -(\mathbf{f}_p)^\top \dot{\mathbf{q}}_p \quad (17)$$

while the local definition of flux density per unit surface can be replaced by the partial fluxes flowing out of one or the other of the cellular sub-boundaries. Specifically, the energy fluxes

$$J^- = -(\mathbf{f}_p^-)^\top \dot{\mathbf{q}}_p^-, \qquad J^+ = -(\mathbf{f}_p^+)^\top \dot{\mathbf{q}}_p^+ \quad (18)$$

give the mechanical energy flowing out of the negative and positive microstructural sub-boundaries $\Gamma^-$ and $\Gamma^+$, respectively. Again, in the *free oscillations* regime the $J$-flux flowing out of the cellular microstructural $\Gamma$-boundary is null and $J^- = -J^+$.

In the framework of the wave dynamics, it may be convenient to define a pair of directional flux vectors, collecting columnwise the companion fluxes flowing out of the orthogonal left and bottom segments $\Gamma_l^-$ and $\Gamma_b^-$ (right and top segments $\Gamma_r^+$ and $\Gamma_t^+$) of the negative (positive) sub-boundaries

$$\mathbf{J}^- = \begin{pmatrix} J_l \\ J_b \end{pmatrix} = -\begin{pmatrix} (\mathbf{f}_l^-)^\top \dot{\mathbf{q}}_l^- \\ (\mathbf{f}_b^-)^\top \dot{\mathbf{q}}_b^- \end{pmatrix}, \quad \mathbf{J}^+ = \begin{pmatrix} J_r \\ J_t \end{pmatrix} = -\begin{pmatrix} (\mathbf{f}_r^+)^\top \dot{\mathbf{q}}_r^+ \\ (\mathbf{f}_t^+)^\top \dot{\mathbf{q}}_t^+ \end{pmatrix} \quad (19)$$

which can be regarded as a sort of the Umov-Poynting vectors for the periodic cell of the beam lattice material. The essential difference with respect to homogeneous continua is that the vectors $\boldsymbol{\jmath}^-$ and $\boldsymbol{\jmath}^+$ cannot be considered local quantities (referred to the generic material point). On the contrary, they are referred to a minimal unit of volume (the finite domain of the periodic cell). Attention is focused on the flux vector $\boldsymbol{\jmath}^+$ in the following, and the superscript $()^+$ is omitted for the sake of simplicity.

The Umov-Poynting vector $\boldsymbol{\jmath}$ is known to be a real quantity in elastic media [1]. Indeed, considering the natural free motion of a monochromatic wave with frequency $\omega$ and complex waveform $\boldsymbol{\psi}_a$, the energy flux $\jmath(\tau)$ is the dot product of the time-harmonic vector fields $\mathbf{f}_p^+$ and $\dot{\mathbf{q}}_p^+$, whose amplitudes vary harmonically in time (with the same frequency $\omega$) and multiplies the complex-valued forms $\boldsymbol{\varphi}_p^+$ and $\boldsymbol{\psi}_p^+$. Therefore, if $\mathbf{f}_p^+$ and $\dot{\mathbf{q}}_p^+$ are harmonically varying complex-valued quantities, the corresponding real-valued vector fields are

$$\mathfrak{R}(\mathbf{f}_p^+) = \tfrac{1}{2}\left(\mathbf{f}_p^+ + (\mathbf{f}_p^+)^*\right), \qquad \mathfrak{R}(\dot{\mathbf{q}}_p^+) = \tfrac{1}{2}\left(\dot{\mathbf{q}}_p^+ + (\dot{\mathbf{q}}_p^+)^*\right) \quad (20)$$

and the time-dependent energy flux is the real valued product

$$\mathfrak{R}(\jmath(\tau)) = -\mathfrak{R}(\mathbf{f}_p^+)^\top \mathfrak{R}(\dot{\mathbf{q}}_p^+) \quad (21)$$

where the $()^*$ superscript stands for the complex conjugate.

From the physical viewpoint, it may be convenient to introduce the *mean* energy flux $\bar{\jmath}$, defined as the time-averaged flux over an oscillation cycle of period $2\pi/\omega$

$$\bar{\jmath} = \frac{\omega}{2\pi}\int_0^{\frac{2\pi}{\omega}} \mathfrak{R}(\jmath(\tau))\,d\tau = -\frac{\omega}{2\pi}\int_0^{\frac{2\pi}{\omega}} \mathfrak{R}(\mathbf{f}_p^+)^\top \mathfrak{R}(\dot{\mathbf{q}}_p^+)\,d\tau \quad (22)$$

which is known to be [5, 8, 53]

$$\bar{\jmath} = \tfrac{1}{2}\mathfrak{R}(-(\mathbf{f}_p^+)^\top(\dot{\mathbf{q}}_p^+)^*) = -\tfrac{1}{4}\left((\mathbf{f}_p^+)^\top(\dot{\mathbf{q}}_p^+)^* + (\mathbf{f}_p^+)^\dagger(\dot{\mathbf{q}}_p^+)\right) \quad (23)$$

where the superscript $()^\dagger$ stands for the conjugate transpose.

Moreover, a mean *directional* flux of mechanical energy can be defined, by collecting columnwise the flux components $\bar{\jmath}_r$ and $\bar{\jmath}_t$ flowing out of the boundary segments $\Gamma_r$ and $\Gamma_t$

$$\bar{\boldsymbol{\jmath}} = \begin{pmatrix} \bar{\jmath}_r \\ \bar{\jmath}_t \end{pmatrix} = -\tfrac{1}{4}\begin{pmatrix} (\mathbf{f}_r^+)^\top(\dot{\mathbf{q}}_r^+)^* + (\mathbf{f}_r^+)^\dagger(\dot{\mathbf{q}}_r^+) \\ (\mathbf{f}_t^+)^\top(\dot{\mathbf{q}}_t^+)^* + (\mathbf{f}_t^+)^\dagger(\dot{\mathbf{q}}_t^+) \end{pmatrix} \quad (24)$$

The vector $\bar{\boldsymbol{\jmath}}$ is the time-average of the vector $\boldsymbol{\jmath}^+$ in the equation (19), and is referred to as *Umov-Poynting vector* (for the beam lattice) in the following, for the sake of simplicity.

The passive forces and velocities are linearly depending on the active displacements and velocities through the relations

$$\mathbf{f}_r^+ = -\mathrm{e}^{-\imath\beta_1}\,\mathbf{F}_{la}^-\mathbf{q}_a, \qquad \mathbf{f}_t^+ = -\mathrm{e}^{-\imath\beta_2}\,\mathbf{F}_{ba}^-\mathbf{q}_a \quad (25)$$

$$\dot{\mathbf{q}}_r^+ = \mathrm{e}^{-\imath\beta_1}\,\mathbf{S}_{la}^-\dot{\mathbf{q}}_a, \qquad \dot{\mathbf{q}}_t^+ = \mathrm{e}^{-\imath\beta_2}\,\mathbf{S}_{ba}^-\dot{\mathbf{q}}_a \quad (26)$$

by virtue of the quasi-static condensation and the material periodicity. Consequently, the components of the Umov-Poynting vector $\bar{\boldsymbol{\jmath}}$ can conveniently be expressed as a function of the only active displacements and velocities

$$\bar{\jmath}_r = \tfrac{1}{4}\left(\mathbf{q}_a^\top(\mathbf{F}_{la}^-)^\top(\mathbf{S}_{la}^-)^*\dot{\mathbf{q}}_a^* + \mathbf{q}_a^\dagger(\mathbf{F}_{la}^-)^\dagger\mathbf{S}_{la}^-\dot{\mathbf{q}}_a\right) \quad (27)$$

$$\bar{\jmath}_t = \tfrac{1}{4}\left(\mathbf{q}_a^\top(\mathbf{F}_{ba}^-)^\top(\mathbf{S}_{ba}^-)^*\dot{\mathbf{q}}_a^* + \mathbf{q}_a^\dagger(\mathbf{F}_{ba}^-)^\dagger\mathbf{S}_{ba}^-\dot{\mathbf{q}}_a\right) \quad (28)$$

where the auxiliary complex-valued matrices $\mathbf{F}_{la}^-,\mathbf{F}_{ba}^-,\mathbf{S}_{la}^-,\mathbf{S}_{ba}^-$ are reported in the AppendixA.2.

Making explicit the natural wave solution $\mathbf{q}_a = A\,\boldsymbol{\psi}_a\exp(\imath\omega\tau)$ for the forward propagating wave with frequency $\omega$ and waveform $\boldsymbol{\psi}_a$, the flux components can be expressed in the form

$$\bar{\jmath}_r = \tfrac{AA^*}{4}\imath\omega\left(-\boldsymbol{\psi}_a^\top(\mathbf{F}_{la}^-)^\top(\mathbf{S}_{la}^-)^*\boldsymbol{\psi}_a^* + \boldsymbol{\psi}_a^\dagger(\mathbf{F}_{la}^-)^\dagger\mathbf{S}_{la}^-\boldsymbol{\psi}_a\right) \quad (29)$$

$$\bar{\jmath}_t = \tfrac{AA^*}{4}\imath\omega\left(-\boldsymbol{\psi}_a^\top(\mathbf{F}_{ba}^-)^\top(\mathbf{S}_{ba}^-)^*\boldsymbol{\psi}_a^* + \boldsymbol{\psi}_a^\dagger(\mathbf{F}_{ba}^-)^\dagger\mathbf{S}_{ba}^-\boldsymbol{\psi}_a\right) \quad (30)$$

and can be proved to be real-valued, since the terms between brackets are purely imaginary (see also AppendixB.2). Recalling the significant relation $\boldsymbol{\psi}_a = \mathbf{Q}^{-1}\boldsymbol{\phi}_a$, the flux components can also be expressed as function of the standard eigenvectors

$$\bar{\jmath}_r = \tfrac{AA^*}{4}\imath\left(-\boldsymbol{\phi}_a^\top\mathbf{J}_r\boldsymbol{\phi}_a^* + \boldsymbol{\phi}_a^\dagger\mathbf{J}_r^*\boldsymbol{\phi}_a\right) \quad (31)$$

$$\bar{\jmath}_t = \tfrac{AA^*}{4}\imath\left(-\boldsymbol{\phi}_a^\top\mathbf{J}_t\boldsymbol{\phi}_a^* + \boldsymbol{\phi}_a^\dagger\mathbf{J}_t^*\boldsymbol{\phi}_a\right) \quad (32)$$

where the auxiliary square matrices

$$\mathbf{J}_r = \omega\,\mathbf{Q}^{-\top}(\mathbf{F}_{la}^-)^\top(\mathbf{S}_{la}^-)^*\mathbf{Q}^{-1} \quad (33)$$

$$\mathbf{J}_t = \omega\,\mathbf{Q}^{-\top}(\mathbf{F}_{ba}^-)^\top(\mathbf{S}_{ba}^-)^*\mathbf{Q}^{-1} \quad (34)$$

are complex-valued. Separating the real and imaginary parts of the eigenvectors $\boldsymbol{\phi}_a = \boldsymbol{\chi} + \imath\boldsymbol{\gamma}$, the formulas (31),(32) read

$$\bar{\jmath}_r = \tfrac{AA^*}{2}\left(\boldsymbol{\chi}^\top\mathfrak{I}(\mathbf{J}_r)\boldsymbol{\chi} + \boldsymbol{\gamma}^\top\mathfrak{I}(\mathbf{J}_r)\boldsymbol{\gamma} - \boldsymbol{\chi}^\top\mathfrak{R}(\mathbf{J}_r)\boldsymbol{\gamma} + \boldsymbol{\gamma}^\top\mathfrak{R}(\mathbf{J}_r)\boldsymbol{\chi}\right) \quad (35)$$

$$\bar{\jmath}_t = \tfrac{AA^*}{2}\left(\boldsymbol{\chi}^\top\mathfrak{I}(\mathbf{J}_t)\boldsymbol{\chi} + \boldsymbol{\gamma}^\top\mathfrak{I}(\mathbf{J}_t)\boldsymbol{\gamma} - \boldsymbol{\chi}^\top\mathfrak{R}(\mathbf{J}_t)\boldsymbol{\gamma} + \boldsymbol{\gamma}^\top\mathfrak{R}(\mathbf{J}_t)\boldsymbol{\chi}\right) \quad (36)$$

where the real and imaginary parts of the auxiliary matrices are

$$\mathfrak{R}(\mathbf{J}_r) = \omega\mathbf{Q}^{-\top}\left(\mathfrak{I}(\mathbf{F}_{la}^-)^\top\mathfrak{I}(\mathbf{S}_{la}^-) + \mathfrak{R}(\mathbf{F}_{la}^-)^\top\mathfrak{R}(\mathbf{S}_{la}^-)\right)\mathbf{Q}^{-1} \quad (37)$$

$$\mathfrak{R}(\mathbf{J}_t) = \omega\mathbf{Q}^{-\top}\left(\mathfrak{I}(\mathbf{F}_{ba}^-)^\top\mathfrak{I}(\mathbf{S}_{ba}^-) + \mathfrak{R}(\mathbf{F}_{ba}^-)^\top\mathfrak{R}(\mathbf{S}_{ba}^-)\right)\mathbf{Q}^{-1} \quad (38)$$

$$\mathfrak{I}(\mathbf{J}_r) = \omega\mathbf{Q}^{-\top}\left(\mathfrak{I}(\mathbf{F}_{la}^-)^\top\mathfrak{R}(\mathbf{S}_{la}^-) - \mathfrak{R}(\mathbf{F}_{la}^-)^\top\mathfrak{I}(\mathbf{S}_{la}^-)\right)\mathbf{Q}^{-1} \quad (39)$$

$$\mathfrak{I}(\mathbf{J}_t) = \omega\mathbf{Q}^{-\top}\left(\mathfrak{I}(\mathbf{F}_{ba}^-)^\top\mathfrak{R}(\mathbf{S}_{ba}^-) - \mathfrak{R}(\mathbf{F}_{ba}^-)^\top\mathfrak{I}(\mathbf{S}_{ba}^-)\right)\mathbf{Q}^{-1} \quad (40)$$

An equivalent alternative formulation of the Umov-Poynting vector is reported in AppendixB.2. As final remark, the mean complementary flux $\bar{\boldsymbol{\jmath}}^- = (\bar{\jmath}_l, \bar{\jmath}_b)$ obeys to the balance condition $(\bar{\jmath}_l + \bar{\jmath}_b) = -(\bar{\jmath}_r + \bar{\jmath}_t)$, according to the free oscillation assumption.

### 2.4. Energy velocity

In classical homogeneous media, which are conservative and nondispersive, the velocity $\mathbf{c}_e$ of the mechanical energy transported by a monocromatic planar wave is a harmonically-varying vector field. According to a well-established definition, the energy velocity vector is determined by the ratio

$$\mathbf{c}_e = \frac{\bar{\mathbf{s}}}{\bar{e}} \quad (41)$$

where $\bar{\mathbf{s}}$ and $\bar{e}$ are the mean Umov-Poynting vector and the mean energy density, averaged over an oscillation period. Within this context, the energy velocity $\mathbf{c}_e$ coincides the group velocity $\mathbf{c}_g$, and its magnitude is also equal to the amplitude $c_p = \omega/\|\mathbf{b}\|$ of the phase velocity. The relation (41) holds also in the presence of dispersion due to heterogeneities, and the energy velocity is found to be equal to the group velocity [17, 18], which in general differs from the phase velocity [51].

In the context of periodic materials described by crystal lattice and beam lattice models, a formally analogous definition can be adopted to describe the velocity vector $\mathbf{v}_e$ related to the mean directional flux of mechanical energy flowing out of the microstructural boundary of the periodic cell

$$\mathbf{v}_e = \frac{\bar{\mathbf{J}}}{\bar{\mathcal{E}}} \tag{42}$$

where the denominator is the mean total energy $\bar{\mathcal{E}}$ stored in the periodic cell, averaged over an oscillation period. According to the balance law of the mechanical energy in the absence of dissipation, the mean total energy can be expressed

$$\bar{\mathcal{E}} = 2\bar{\mathcal{K}} \tag{43}$$

where $\bar{\mathcal{K}}$ is the average kinetic energy, which is also equal to the average potential energy $\bar{\mathcal{U}}_a$ stored in the active degrees-of-freedom (see the AppendixC).

Assuming again the natural wave solution $\mathbf{q}_a = A\,\boldsymbol{\psi}_a \exp(\iota\omega\tau)$ for the forward propagating wave with frequency $\omega$ and waveform $\boldsymbol{\psi}_a$, the average kinetic energy reads

$$\bar{\mathcal{K}} = \tfrac{AA^*}{8}\,\omega^2 \left( \boldsymbol{\psi}_a^\top \mathbf{M}_a \boldsymbol{\psi}_a^* + \boldsymbol{\psi}_a^\dagger \mathbf{M}_a \boldsymbol{\psi}_a \right) \tag{44}$$

or, in terms of standard eigenvectors

$$\bar{\mathcal{K}} = \tfrac{AA^*}{8}\,\omega^2 \left( \boldsymbol{\phi}_a^\top \boldsymbol{\phi}_a^* + \boldsymbol{\phi}_a^\dagger \boldsymbol{\phi}_a \right) \tag{45}$$

Therefore, employing the equations (29),(30) for the numerator and the equations (43),(44) for the denominator, the components of the velocity vector $\mathbf{v}_e = (v_r, v_t)$ read

$$v_r = \frac{\iota\left(-\boldsymbol{\psi}_a^\top (\mathbf{F}_{la}^-)^\top (\mathbf{S}_{la}^-)^* \boldsymbol{\psi}_a^* + \boldsymbol{\psi}_a^\dagger (\mathbf{F}_{la}^-)^\dagger \mathbf{S}_{la}^- \boldsymbol{\psi}_a\right)}{\omega\left(\boldsymbol{\psi}_a^\top \mathbf{M}_a \boldsymbol{\psi}_a^* + \boldsymbol{\psi}_a^\dagger \mathbf{M}_a \boldsymbol{\psi}_a\right)} \tag{46}$$

$$v_t = \frac{\iota\left(-\boldsymbol{\psi}_a^\top (\mathbf{F}_{ba}^-)^\top (\mathbf{S}_{ba}^-)^* \boldsymbol{\psi}_a^* + \boldsymbol{\psi}_a^\dagger (\mathbf{F}_{ba}^-)^\dagger \mathbf{S}_{ba}^- \boldsymbol{\psi}_a\right)}{\omega\left(\boldsymbol{\psi}_a^\top \mathbf{M}_a \boldsymbol{\psi}_a^* + \boldsymbol{\psi}_a^\dagger \mathbf{M}_a \boldsymbol{\psi}_a\right)} \tag{47}$$

where it can be verified that $\boldsymbol{\psi}_a^\top \mathbf{M}_a \boldsymbol{\psi}_a^* = \boldsymbol{\psi}_a^\dagger \mathbf{M}_a \boldsymbol{\psi}_a$, according to the properties of the eigenvectors of hermitian matrices. Therefore, the denominator can be expressed as $2\omega\,\boldsymbol{\psi}_a^\dagger \mathbf{M}_a \boldsymbol{\psi}_a$.

### 2.5. Group velocity

The group velocity $\mathbf{c}_g = (c_{g1}, c_{g2})$ is physically representative of the velocity with which the wave modulation or envelope propagates. For the natural wave $\mathbf{q}_a = A\,\boldsymbol{\psi}_a \exp(\iota\omega\tau)$ propagating wave with frequency $\omega(\mathbf{b})$, the classic definition is

$$\mathbf{c}_g = \nabla_\mathbf{b} \omega \tag{48}$$

where $\nabla_\mathbf{b}$ stands for the gradient in the transformed space of the wavevector $\mathbf{b}$. Therefore, according to the differential geometry, the $\mathbf{c}_g$-vector is collinear to the unit vector orthogonal to the iso-frequency curves of the dispersion surface $\omega(\mathbf{b})$.

From the methodological viewpoint, the components of the group velocity can be determined from the characteristic function $F(\lambda, \mathbf{p}, \mathbf{b})$, according to the implicit function theorem, as

$$c_{gi} = \frac{\partial\omega}{\partial\beta_i} = \frac{2}{\omega}\frac{\partial\lambda}{\partial\beta_i} = -\frac{2}{\omega}\frac{\partial_{\beta_i} F(\lambda, \mathbf{p}, \mathbf{b})}{\partial_\lambda F(\lambda, \mathbf{p}, \mathbf{b})} \tag{49}$$

where $\partial_\lambda$ and $\partial_{\beta_i}$ stand for the $\lambda$-derivative and the $\beta_i$-derivative, respectively, with $\beta_i$ component of $\mathbf{b} = (\beta_1, \beta_2)$ and $i = 1, 2$. An analytical formula for the asymptotic approximation of the derivatives in the expression (49) can be found in [57].

An alternative method to determine the group velocity can be founded on the $\beta_i$-derivative of the condensed equations of motion (3). This approach leads to the ratio between two quadratic functions of the complex-valued waveforms $\boldsymbol{\psi}_a$, and employs the $\beta_i$-derivative of the hermitian stiffness matrix $\mathbf{K}_a$

$$c_{gi} = \frac{1}{2\omega} \frac{\boldsymbol{\psi}_a^\dagger \partial_{\beta_i} \mathbf{K}_a \boldsymbol{\psi}_a}{\boldsymbol{\psi}_a^\dagger \mathbf{M} \boldsymbol{\psi}_a} \tag{50}$$

where the numerator is real-valued for the properties of the complex quadratic forms with a hermitian matrix. A proper demonstration of this formula is extensively reported in the AppendixD.1. The equation (50) can alternatively be carried out by adopting a multiparameter perturbation technique, which employs the wavevector $\mathbf{b}$ as perturbation parameter [58].

Finally, a third method to determine the group velocity can be based on the $\beta_i$-derivative of the non-condensed equations of motion (2). This approach differs from the previous one for the employment of the complex-valued passive waveforms $\boldsymbol{\psi}_p$

$$c_{gi} = \frac{1}{2\omega} \frac{\boldsymbol{\varphi}_p^\dagger \partial_{\beta_i} \boldsymbol{\psi}_p - \boldsymbol{\psi}_p^\dagger \partial_{\beta_i} \boldsymbol{\varphi}_p}{\boldsymbol{\psi}_a^\dagger \mathbf{M} \boldsymbol{\psi}_a} \tag{51}$$

where $\boldsymbol{\varphi}_p = (\boldsymbol{\varphi}_p^-, \boldsymbol{\varphi}_p^+)$ are the waveforms of the passive forces, which can be related to the passive waveforms through the relation $\boldsymbol{\varphi}_p^- = \mathbf{F}_{pp}^- \boldsymbol{\psi}_p^-$ and $\boldsymbol{\varphi}_p^+ = -\mathbf{L}\boldsymbol{\varphi}_p^- = -\mathbf{L}\mathbf{F}_{pp}^- \boldsymbol{\psi}_p^-$. A proper demonstration of this formula is extensively reported in the AppendixD.2, in analogy to the procedure proposed in [52]. An alternative demonstration can be carried out by adopting a multiparameter perturbation technique [58]. It is worth noting that the group velocity turns out to be equal to the energy velocities (46),(47), consistently with the findings of [10, 52]. The details of the demonstration are reported in the AppendixD.3.

### 3. Tetrachiral material

The two-dimensional geometric pattern of the tetrachiral material is characterized by square cells realizing a regular, periodic tessellation of the infinite euclidean plane (Figure 4a). The mechanical behaviour of each elementary cell is determined by a distinctive centro-symmetric microstructure, including a central circular ring connected to four tangent ligaments organized according to a chiral topology (Figure 4b). The particular architecture of the cellular microstructure determines strong properties of anisotropy in the macroscopic response of the material. Furthermore, the marked auxeticity characterizing certain stretching directions makes the tetrachiral material a promising candidate for advanced engineering applications [59–64].

The tetrachiral material may represent a suited benchmark to study the polarization of acoustic waves and the transport of mechanical energy in beam lattice models. Indeed, a few mechanical hypotheses about the cell microstructure are sufficient to formulate a low-dimensional but representative Lagrangian

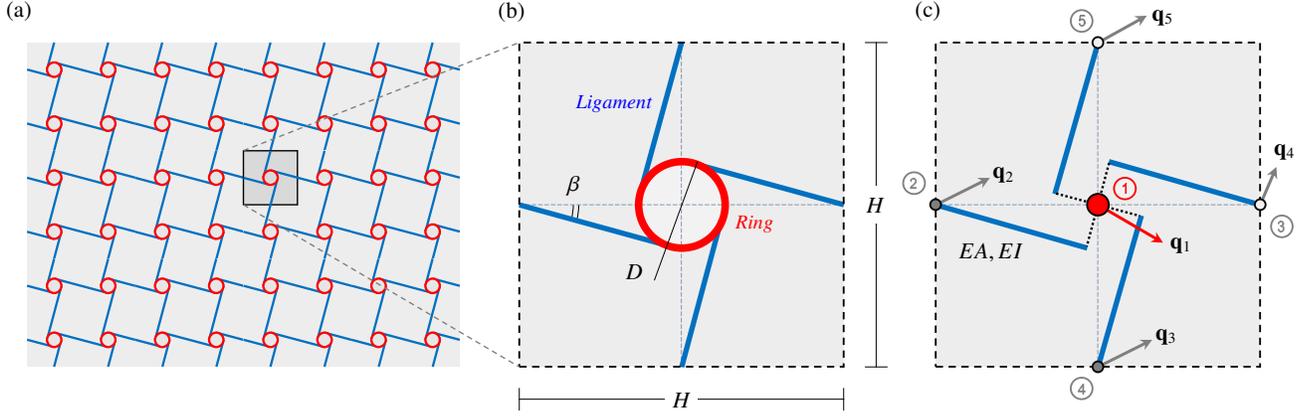

Figure 4: Tetrachiral metamaterial (a) repetitive planar pattern, (b) centro-symmetric microstructure of the periodic cell, (c) beam lattice model (black dotted lines represent rigid-end offsets connecting the beam-ring connections with the ring centroid

model (Figure 4c). Specifically, the circular ring is assumed heavy and sufficiently stiff to be modeled as a massive rigid body. The ring mass $M$ and rotational inertia $J$ can be freely assigned, by independently setting the mean diameter $D$, the annular width and the mass density. The varied configuration of the rigid body is described by the three planar displacements, namely the in-plane translations $U_1, V_1$ and rotation $\theta_1$ of the configurational node ① located at the ring centroid. The four identical ligaments are supposed sufficiently light and flexible to be described by linear, extensible, unshearable and massless beams, with elastic and geometric properties defining the extensional $EA$ and flexural rigidity $EI$. The natural length of the beams is $L = H\cos\beta$, where $H$ is the cell side length and $\beta = \arcsin(D/H)$ is the ligament inclination angle (*chirality angle*) with respect to the mesh lines connecting all the ring centres. The ring-beam connections are considered perfectly-rigid joints at the four points of tangency. Therefore, the planar displacements of the ring-beam joints are entirely dependent on the ring centroid motion, by virtue of the rigid body assumption. Due to the geometric periodicity, the cell boundary crosses the midspan of each ligament. Consequently, the varied configuration of the $i$-th beam is fully described by the rigid motion of the ligament-ring connection, at the one end, and the three planar displacements ($U_i, V_i, \theta_i$) of the $i$-th configurational node ($i = $②,...,⑤) located at the cell boundary, at the other end.

Denoting $\Omega_c$ a known circular frequency serving as auxiliary dimensional reference, the inertial, elastic and geometric properties of the cell microstructure are described by the minimal set $\mathbf{p}$ of independent nondimensional parameters

$$\delta = \frac{D}{H}, \quad \varrho^2 = \frac{I}{AL^2}, \quad \chi^2 = \frac{J}{MH^2}, \quad \omega_c^2 = \frac{EA}{MH\Omega_c^2} \quad (52)$$

where $\delta$ expresses the spatial density of the rings, measuring also the material mass density at the macroscopic scale. Furthermore, the parameter $\delta$ is an indirect measure of the material chirality through the immediate relation $\delta = \sin\beta$ with the chirality angle $\beta$. The inverse of the nondimensional radius of gyration $\varrho$ accounts for the beam slenderness. The parameter $\chi^2$ describes the rotational-to-translational mass ratio of the rings and, for homogeneous mass distribution, essentially depends on the ring thickness. Finally, the parameter $\omega_c$ describes a nondimensional normalization frequency, which can be assumed to be unitary in the following, without loss of generality.

### 3.1. Equation of motion

According to the mechanical assumptions, the linear dynamics of the periodic cell is governed by a multi-degrees-of-freedom model, referred to six configuration nodes. The actual configuration of the $i$-th node is described by the nondimensional displacement vector $\mathbf{q}_i = (u_i, v_i, \theta_i)$, where the variables

$$u_i = \frac{U_i}{H}, \qquad v_i = \frac{V_i}{H} \quad (53)$$

are the nondimensional displacements in the material plane and $\theta_i$ is the out-of-plane rotation. The cell configuration is described by the displacement vector $\mathbf{q} = (\mathbf{q}_1, ..., \mathbf{q}_5)$, with $N = 15$.

Employing the direct stiffness method to impose the equilibrium, the undamped free response of the Lagrangian model is governed by ordinary differential equations of motion that assume the general form of equation (1). The configuration vector can be partitioned in the form $\mathbf{q} = (\mathbf{q}_a, \mathbf{q}_p)$, where

- the *active* displacement vector $\mathbf{q}_a$ coincides with $\mathbf{q}_1$, and collects the degrees-of-freedom of the central node at the ring centroid (*internal node*)

- the *passive* displacement vector $\mathbf{q}_p = (\mathbf{q}_2, ..., \mathbf{q}_5)$ collects the degrees-of-freedom of the midspan nodes of the four ligaments (*external nodes*).

with $N_a = 3$ and $N_p = 12$. According to the displacement partition, the equations of motion can be expressed in the partitioned form of equation (2). The mass and stiffness matrices $\mathbf{M}$ and $\mathbf{K}$ of the tetrachiral material are reported in the AppendixE.1. $\mathcal{B} = [-\pi, \pi] \times [-\pi, \pi]$.

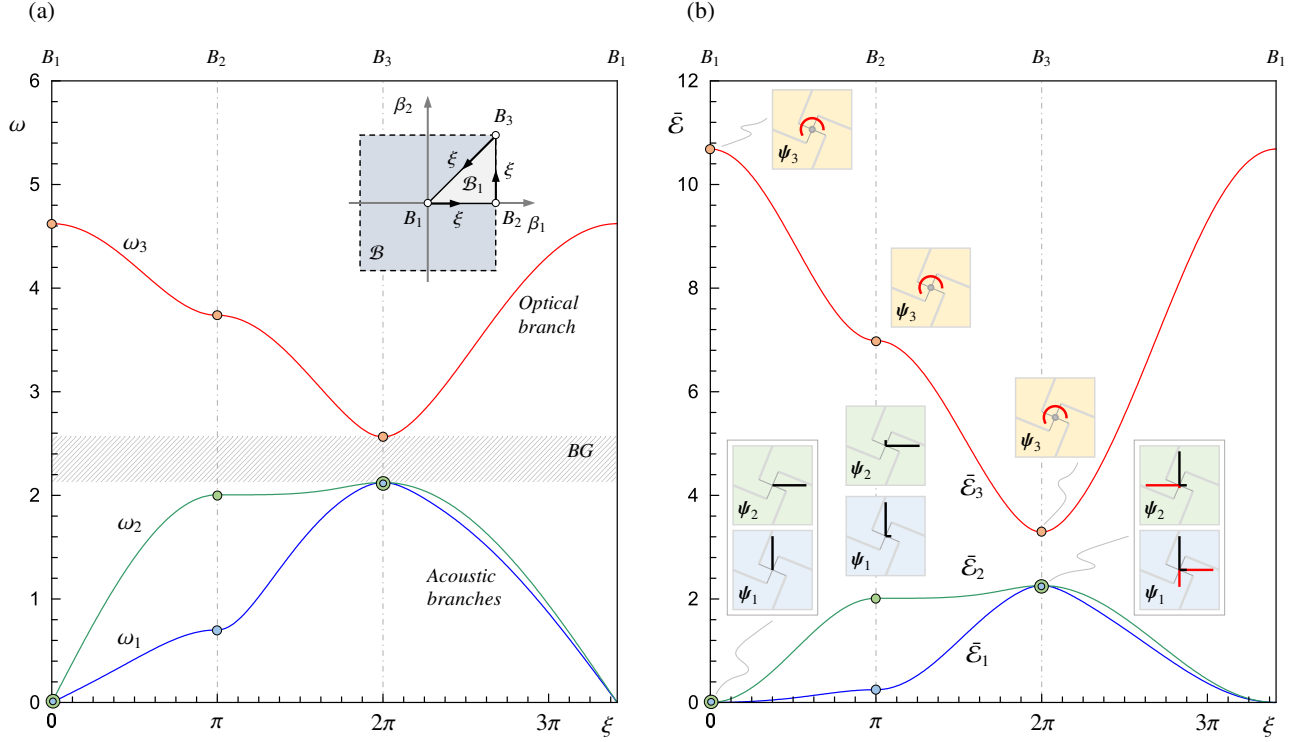

Figure 5: Dispersion spectrum of the tetrachiral material: (a) frequency spectrum, (b) mechanical energy spectrum with selected complex-valued waveforms.

The microstructural boundary $\Gamma$ of the periodic cell can be decomposed into the negative sub-boundary $\Gamma^- = \Gamma_l^- \cup \Gamma_b^+$ and the positive sub-boundary $\Gamma^+ = \Gamma_r^+ \cup \Gamma_t^+$. In the specific case of the tetrachiral material, each boundary segment includes a single external node of the cell microstructure. Consequently, the passive displacement vector is readily decomposed in the form $\mathbf{q}_p = (\mathbf{q}_p^-, \mathbf{q}_p^+)$. Furthermore, the passive displacement sub-vectors are immediately associated $\mathbf{q}_p^- = (\mathbf{q}_l^-, \mathbf{q}_b^-) = (\mathbf{q}_2, \mathbf{q}_4)$ and $\mathbf{q}_p^+ = (\mathbf{q}_r^+, \mathbf{q}_t^+) = (\mathbf{q}_3, \mathbf{q}_5)$, with $N_l = N_r = 3$ and $N_b = N_t = 3$.

Finally, imposing the quasi-periodicity conditions and applying the quasi-static condensation on the *passive* displacements leads to a condensed equation of motion that can be arranged in the form of equation (3). The **b**-dependent condensed matrix $\mathbf{K}_a$ of the tetrachiral material is reported in the AppendixE.2. In summary, the free propagation of elastic waves in the tetrachiral material is governed by a low-dimension (condensed) beam lattice model, in which the cell configuration is fully described by the three active degrees-of-freedom of the central rigid ring.

### 3.2. Dispersion spectrum

Imposing the harmonic solution for forward propagating waves $\mathbf{q}_a = \boldsymbol{\psi}_a \exp(\iota \omega \tau)$ in the equations of motion, the linear eigenproblem governing the wave propagation in the tetrachiral material assumes the classic form of equation (4) or, equivalently, the standard form of equation (5). The eigenproblem solution gives three **b**-dependent eigenpairs $(\omega_i, \boldsymbol{\psi}_{ai})$, including the wave frequency $\omega_i$ and the corresponding complex-valued waveform $\boldsymbol{\psi}_{ai}$ ($i = 1, 2, 3$). Fixed a certain set **p** of mechanical parameters, the dispersion relations $\omega(\mathbf{b})$ of the tetrachiral material is fully characterized by the three-valued dispersion function $\omega(\xi)$, where the curvilinear abscissa $\xi$ spans the closed boundary of the triangular domain $\mathcal{B}_1$ (identified by the vertices $B_1, B_2, B_3$ in Figure 5a), contained in the first Brillouin zone $\mathcal{B}$.

The dispersion spectrum is featured by two acoustic branches related to the lower (first and second) frequencies $\omega_1$ and $\omega_2$, and a single optical branch related to the highest (third) frequency $\omega_3$. Figure 5a illustrates the dispersion spectrum for a particular parameter set $\mathbf{p}_*$ corresponding to a typical cell microstructure, characterized by low material density and low chirality ($\delta = \sin\beta = 1/10$), highly-flexible ligaments ($\varrho = 1/10$) and thin rings ($\chi = 1/9$).

The two acoustic branches attain null frequency values ($\omega_1 = \omega_2 = 0$) in the limit of infinitely *long wavelengths*, represented by the vertex $B_1$ (identified by the abscissa $\xi = 0$ and corresponding to vanishing wavenumbers, or formally $|\mathbf{b}| \to \mathbf{0}$). Starting from this limit, the acoustic frequencies grow up almost linearly, with different slopes, for decreasing wavelengths (increasing wavenumbers). The initial linearity of the frequency growth tends to decay with the increasing distance from the vertex $B_1$. Finally, the dispersion relation $\omega_i(\xi)$ becomes strongly nonlinear in the proximity of the limits of *short wavelengths* represented by the vertices $B_2$ (identified by the abscissa $\xi = \pi$ and pointed by the wavevector $\mathbf{b} = (\pi, 0)$) and $B_3$ (identified by the abscissa $\xi = 2\pi$ and pointed by the wavevector $\mathbf{b} = (\pi, \pi)$). On the contrary, the optical branch attains a not-null, high frequency value ($\omega_3 \simeq 4.6$) in the limit of long wavelengths. Therefore, the optical frequency decreases with a marked nonlinear dependence on the increasing wavenumbers. This persistent qualitative scenario can be verified to characterize the dispersion spectrum of the tetrachiral material over large ranges of the mechanical parameters (wide parametric analyses are reported in [65]). As complementary remark, it is worth noting that the full band gap *BG* exists between the maximum of the acoustic low-frequency branches and the minimum of the optical high-frequency branch. The parametric conditions for the opening/closing of this band gap can be determined analytically, by virtue of perturbation methods for the asymptotic solution of the eigenproblem [58].

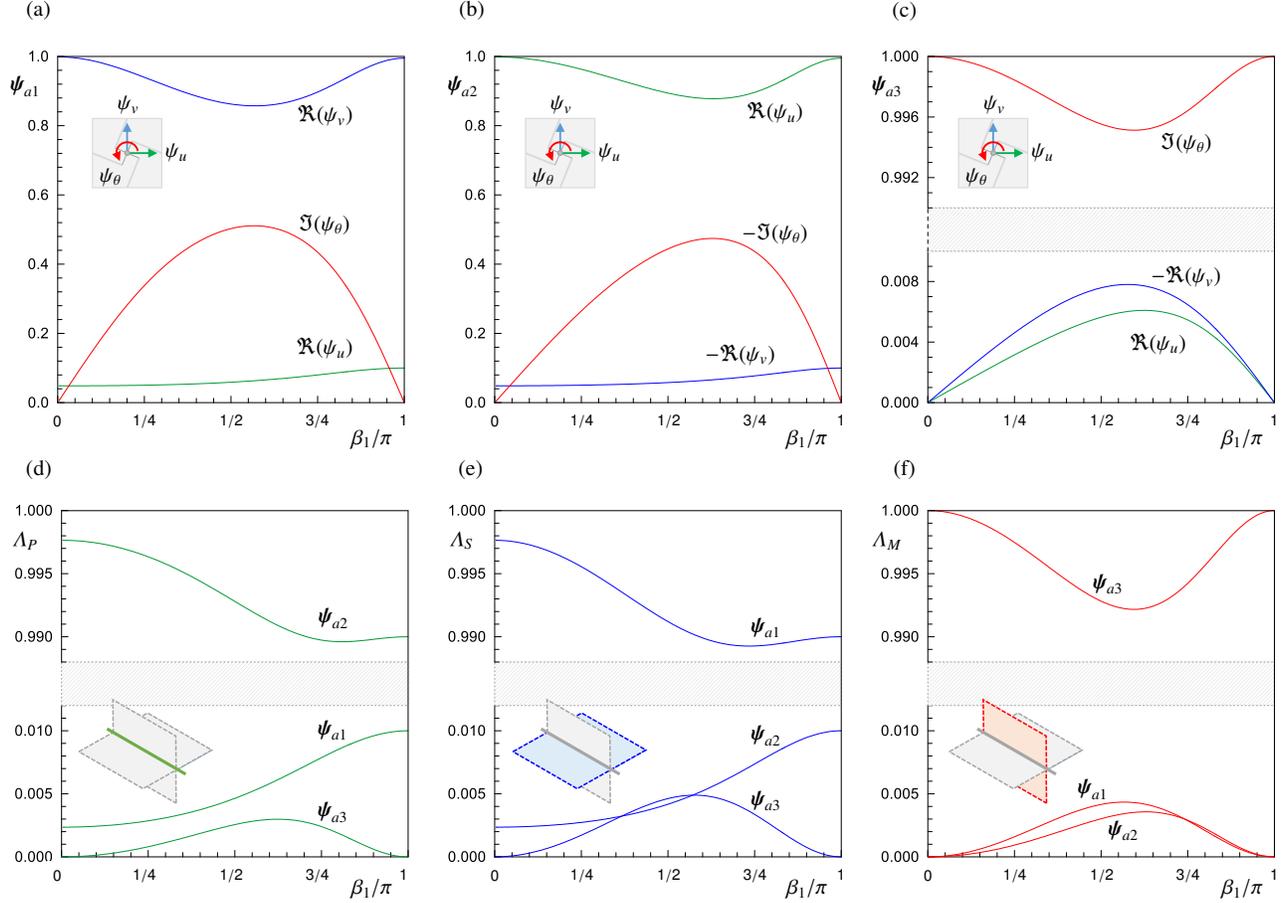

Figure 6: Waveforms $\boldsymbol{\psi}_a = (\psi_u, \psi_v, \psi_\theta)$ and polarization factors versus the wavenumber $\beta_1$ (horizontally propagating waves) for the tetrachiral material with parameter set $\mathbf{p}^*$ ($\delta = \sin\beta = 1/10, \varrho = 1/10, \chi = 1/9$): (a) components of the first waveform $\boldsymbol{\psi}_{a1}$, (b) components of second waveform $\boldsymbol{\psi}_{a2}$, (c) components of third waveform $\boldsymbol{\psi}_{a3}$, (d) complementary compression factor $\Lambda_P$, (e) polarization shear factor $\Lambda_S$, (f) polarization moment factor $\Lambda_M$.

The mechanical energy spectrum can be determined by evaluating the mean total energy $\bar{\mathcal{E}}$ transferred by the forward propagating wave $\mathbf{q}_a = A\boldsymbol{\psi}_a \exp(\iota\omega\tau)$ according to the equation (43). Figure 5b illustrates the energy spectrum for the particular parameter set $\mathbf{p}_*$. The oscillation amplitude $A$ is assumed unitary and the waveform $\boldsymbol{\psi}_a$ is mass-normalized, for the sake of simplicity. Coherently with the frequency spectrum, the energy spectrum possesses two acoustic branches related to the lower frequency waves, determining small energy transfers $\bar{\mathcal{E}}_1$ and $\bar{\mathcal{E}}_2$, and a single optical branch related to the highest frequency wave, determining large energy transfers $\bar{\mathcal{E}}_3$. This evidence highlights that – considering equal oscillation amplitudes – the energy transferred by the high-frequency *optical wave* (rigorously, the harmonic wave associated with the frequency of the optical branch) is larger than the energy transferred by the low-frequency *acoustic waves* (rigorously, the harmonic waves associated with the frequencies of the acoustic branches). From a different viewpoint, it could be said that generating a high-frequency optical wave in the tetrachiral material requires larger energy than generating low-frequency acoustic waves propagating with the same oscillation amplitude. The difference in the energy transferred by different waves has a minimum for the limit of short and equal wavelengths (or large identical wavenumbers). This limit corresponds to the vertex $B_3$ (abscissa $\xi = 2\pi$), where the maximum of the acoustic energy branches co-exists with the minimum of the optical energy branch. The existence of the energy band gap obeys to the same conditions valid for the frequency band gap.

Due to the absence of dissipation, the total energy $\bar{\mathcal{E}}$ is twice the elastic energy $\bar{\mathcal{U}}_a$ stored in the cell microstructure (see also the AppendixC). Therefore, it may be convenient to enrich the discussion about the energy spectrum with some considerations about the waveforms related to each spectral branch. Indeed, each waveform can be regarded as the elastic displacement vector caused by the static application of inertial forces proportional to the waveform itself. The framed boxes in Figure 5b show a few samples of the waveforms $\boldsymbol{\psi}_{ai} = (\psi_u, \psi_v, \psi_\theta)$. More specifically, the triplet of waveforms ($i = 1, 2, 3$) related to each of the three $\mathcal{B}_1$-vertices are reported, while the waveform variation under continuous $\xi$-increment is treated with more depth in the next paragraph. For each component of the $\boldsymbol{\psi}_{ai}$-vector, or *wavecomponent*, the real part (black) is distinguished from the imaginary part (red). As first remark, null elastic energy is associated to the perfectly polarized waveforms $\boldsymbol{\psi}_{a1}$ and $\boldsymbol{\psi}_{a2}$ of the acoustic waves with infinitely long wavelengths (vertex $B_1$), corresponding to rigid (null-frequency) translations in the material plane. The maximum elastic energy for the acoustic waves, reached in the limit of short wavelengths (vertex $B_3$), is instead associated to an iso-frequency pair of hybrid waveforms $\boldsymbol{\psi}_{a1}$ and $\boldsymbol{\psi}_{a2}$, characterized by a comparable participation of the in-plane wavecomponents $\psi_u$ and $\psi_v$. On the contrary, the optical branch is characterized by a perfectly polarized waveform $\boldsymbol{\psi}_{a3}$, in which the out-of-plane rotation $\psi_\theta$ is the only not null wavecomponent and no hybridizations occur in the limits of long and short wavelengths (vertices $B_1$ and $B_3$).

### 3.3. Wave polarization

The variation of the wave frequency along the closed boundary of the triangular zone $\mathcal{B}_1$, described by the dispersion function $\omega(\xi)$, is accompanied by the $\xi$-dependent variation of the corresponding waveform $\boldsymbol{\psi}_a = (\psi_u, \psi_v, \psi_\theta)$. Adopting a suited self-normalization for each waveform ($\boldsymbol{\psi}_a^\dagger \boldsymbol{\psi}_a = 1$), the components of the three vector functions $\boldsymbol{\psi}_{a1}(\xi), \boldsymbol{\psi}_{a2}(\xi), \boldsymbol{\psi}_{a3}(\xi)$ are reported for the first side of the triangular boundary ($0 \leq \xi \leq \pi$), corresponding to horizontally propagating waves (see Figures 6a,b,c). Considering the first frequency $\omega_1$, the corresponding waveform $\boldsymbol{\psi}_{a1}$ is dominated by the vertical wavecomponent $\psi_v$ (Figures 6a). This qualitative remark is accurately quantified by the corresponding shear polarization factor $\Lambda_S$, which is approximately unitary along the whole boundary side (see the function $\Lambda_S(\xi) \simeq 1$ for $\boldsymbol{\psi}_{a1}$ in Figures 6e). It may be worth noting that the shear polarization is not perfect (note $\Lambda_S(\xi) < 1$), due to a minimal but not negligible participation of the horizontal and rotational wavecomponents. This defect in the shear polarization of the first waveform $\boldsymbol{\psi}_{a1}$ is caused by the low chirality of the tetrachiral microstructure, and is efficiently quantified by the small values attained by the moment polarization factor $\Lambda_M$ and the complementary compression factor $\Lambda_P$ (note $\Lambda_M(\xi) \simeq 0$ and $\Lambda_P(\xi) \simeq 0$ for $\boldsymbol{\psi}_{a1}$ in Figures 6d,f).

A similar qualitative behaviour can be recognized for the second frequency $\omega_2$, which corresponds to a waveform $\boldsymbol{\psi}_{a2}$ dominated by the horizontal wavecomponent $\psi_u$ (Figures 6b). The complementary compression factor $\Lambda_P$ accurately quantifies this behaviour by attaining values close to unity along the whole boundary side (see the function $\Lambda_P(\xi) \simeq 1$ for $\boldsymbol{\psi}_{a2}$ in Figure 6d). Finally, the waveform $\boldsymbol{\psi}_{a3}$ associated to the third frequency $\omega_3$ is dominated by the rotational wavecomponent $\psi_\theta$ (Figures 6c). Consistently, the moment polarization factor $\Lambda_M$ assumes quasi-unitary values (see the function $\Lambda_M(\xi) \simeq 1$ for $\boldsymbol{\psi}_{a3}$ in Figures 6f). As minor remark, the waveform $\boldsymbol{\psi}_{a3}$ is perfectly moment-polarized at the limits of short and long wavelengths (note $\Lambda_M = 1$ for $\boldsymbol{\psi}_{a3}$ at $\xi = 0$ and $\xi = \pi$ in Figures 6f).

### 3.4. Energy flux

The mean directional flux of mechanical energy is described by the Umov-Poynting vector $\bar{\boldsymbol{j}} = (j_r, j_t)$ for the forward propagating waves $\mathbf{q}_a = \boldsymbol{\psi}_a \exp(\iota \omega \tau)$, evaluated according to the formulas (29),(30) for unitary amplitude $A$ and self-normalized waveform $\boldsymbol{\psi}_a$. The smooth functions of the mean energy fluxes $j_r$ and $j_t$ flowing out of the right and top microstructural boundaries are reported for horizontally propagating waves (functions $j_r(\beta_1)$ and $j_t(\beta_1)$ for $0 \leq \beta_1 \leq \pi$ in Figures 7a,d), vertically propagating waves (functions $j_r(\beta_2)$ and $j_t(\beta_2)$ for $0 \leq \beta_2 \leq \pi$ in Figures 7b,d), and diagonally propagating waves (functions $j_r(\beta_3)$ and $j_t(\beta_3)$ for $0 \leq \beta_3 = (\beta_1^2 + \beta_2^2)^{1/2} \leq \sqrt{2}\pi$ in Figures 7c,f).

Focusing first on the energy flux $\bar{j}_r$ for horizontal waves (see the functions $\bar{j}_r(\beta_1)$ in Figure 7a), it can be immediately remarked that vanishing fluxes are systematically related to the limits of long and short wavelengths (note $\bar{j}_r = \bar{j}_t = 0$ at $\beta_1 = 0, \pi$ for all the waveforms). From the physical viewpoint, this remark states that no energy transport can occur in these wavelength limit conditions. The absolute value of all the energy fluxes grows up with increasing distances from the limits of the wavelength range. In particular, the maximum energy flux is reached in the range of medium wavelengths by the second acoustic wave (see the function $\bar{j}_r(\beta_1)$ of the waveform $\boldsymbol{\psi}_{a2}$ over the range $1/2 < \beta_1/\pi < 3/4$ in Figure 7a). This result can be attributed to the high elastic energy stored in the self-normalized waves with quasi-unitary compression factors (recall $\Lambda_P \simeq 1$ for $\boldsymbol{\psi}_{a2}$ in Figure 6d). Lower energy fluxes are instead associated to the self-normalized waveforms $\boldsymbol{\psi}_{a1}$ and $\boldsymbol{\psi}_{a3}$, which are characterized by large shear and moment polarization factors. It must be highlighted that the higher and lower values attributed to the energy flux are intrinsically related to the particular normalization of the waveforms (the mass-normalization, for instance, would determine different results). As important consequence, the highest spectral energy does not necessarily correspond to the highest energy flux. Indeed, the $\omega_3$-frequency wave with the highest spectral energy $\mathcal{E}_3$ (see Figure 5a) does not determine the maximum energy flux. This result is justified by the nonlinear (quadratic) dependence of the energy flux on the waveform.

The energy fluxes $j_r(\beta_2)$ for vertical waves shows qualitatively similar behaviours (see the functions $\bar{j}_r(\beta_1)$ in Figure 7b). From the quantitative viewpoint, instead, the energy fluxes are much lower than those related to horizontal waves (Figure 7a). This result depends on the propagation direction of the wave is relation with the right microstructural boundary considered. Indeed, the energy fluxes are expected to be higher for horizontally propagating waves (propagation direction orthogonal to the boundary) than for vertically propagating waves (propagation direction parallel to the boundary). Minor qualitative differences can be detected in the energy flux $j_r$ for diagonal waves (see the functions $\bar{j}_r(\beta_3)$ in Figure 7c), since the energy fluxes can attain more than one maximum value (see for instance the flux of the waveform $\boldsymbol{\psi}_{a1}$).

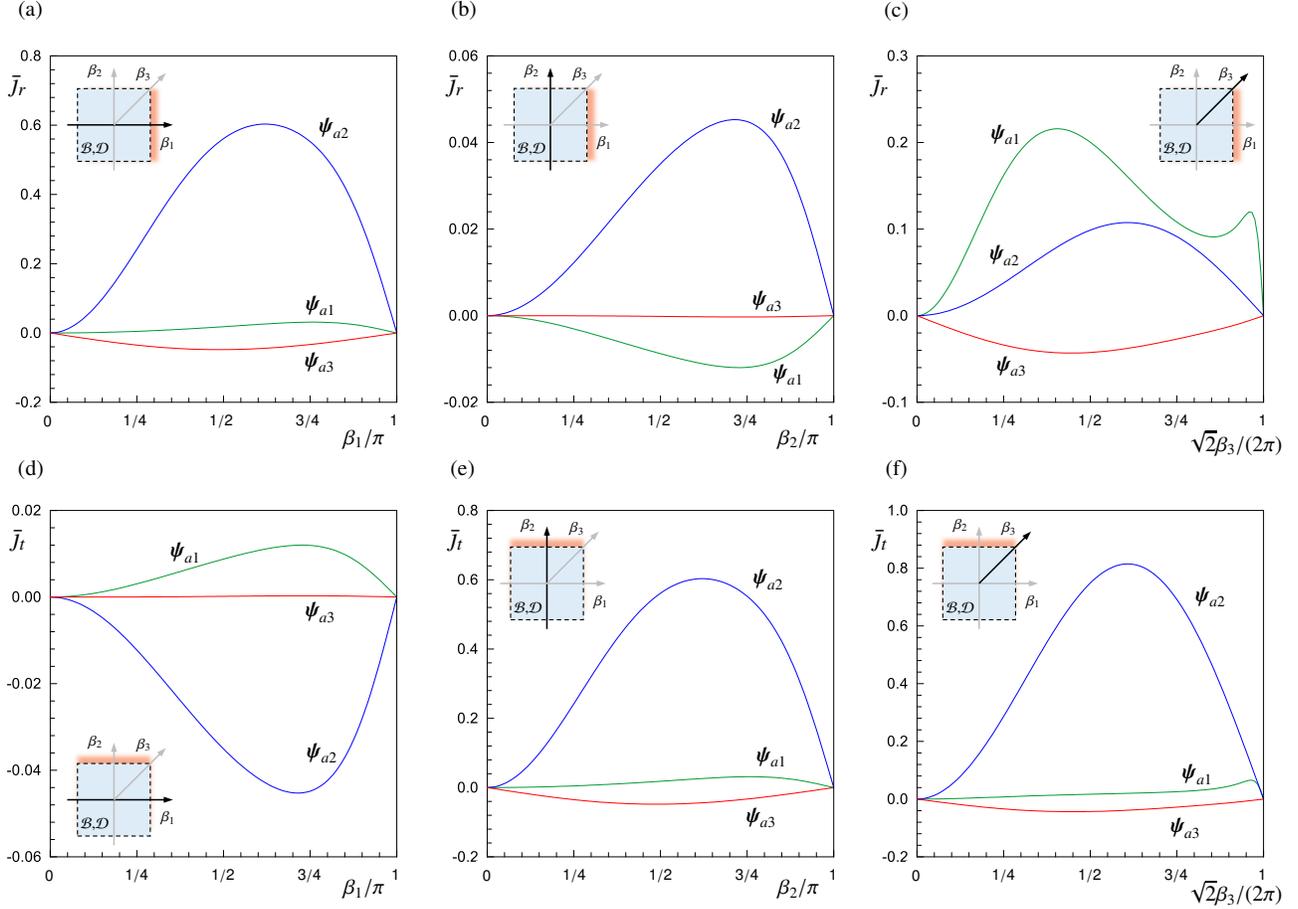

Figure 7: Umov-Poynting vector $\bar{\mathbf{j}} = (\bar{j}_r, \bar{j}_t)$ for the waveforms $\boldsymbol{\psi}_{ai}$ ($i = 1, 2, 3$) of the tetrachiral material with parameter set $\mathbf{p}_*$ ($\delta = \sin\beta = 1/10, \varrho = 1/10, \chi = 1/9$). Energy flux flowing out of the right segment $\Gamma_r$ of the cell boundary: (a) $\bar{j}_r$ versus $\beta_1$ (horizontally propagating waves), (b) $\bar{j}_r$ versus $\beta_2$ (vertically propagating waves), (c) $\bar{j}_r$ versus $\beta_3$ (diagonally propagating waves). Energy flux flowing out of the top segment $\Gamma_t$ of the cell boundary: (d) $\bar{j}_t$ versus $\beta_1$ (horizontally propagating waves), (e) $\bar{j}_t$ versus $\beta_2$ (vertically propagating waves), (f) $\bar{j}_t$ versus $\beta_3$ (diagonally propagating waves).

Focusing on the energy fluxes $\bar{j}_t$ flowing out of the top microstructural boundary (Figures 7d,e,f), the essential findings do not show major qualitative differences. Naturally, quantitative difference can be detected, since the highest energy fluxes are associated to vertical waves (Figure 7e), while lower energy fluxes are associated to horizontal waves (Figure 7d). Further parametric analyses allow to verify that the energy fluxes orthogonal to the propagation direction ($j_t$ for horizontal waves and $j_r$ for vertical waves) tend to null values for a vanishing material chirality ($\delta = \sin\beta \to 0$). Coherently, the energy fluxes related to diagonal waves tend to coincide with each other.

## 3.5. Energy velocity

The velocity $\mathbf{v}_e = (v_r, v_t)$ of mechanical energy flow is determined according to the formulas (46),(47). The smooth functions of the velocity components $v_r$ and $v_t$ are reported for the energy flowing out of the right microstructural boundary (Figure 8a,b,c) and the top microstructural boundary (Figure 8d,e,f). Focus is made on the energy velocity of forward propagating waves traveling in the horizontal direction (functions $v_r(\beta_1)$ and $v_t(\beta_1)$ for $0 \leq \beta_1 \leq \pi$ in Figures 8a,d), vertical direction (functions $v_r(\beta_2)$ and $v_t(\beta_2)$ for $0 \leq \beta_2 \leq \pi$ in Figures 8b,e), diagonal direction (functions $v_r(\beta_3)$ and $v_t(\beta_3)$ for $0 \leq \beta_3 \leq \sqrt{2}\pi$ in Figures 8c,f). All the velocities are found to identically vanish for the limit of short wavelengths (at $\beta_1 = \beta_2 = \pi$). On the contrary, the limit of long wavelengths (at $\beta_1 = \beta_2 = 0$) corresponds to non-null finite values of the energy velocity related to the *acoustic* waveforms $\boldsymbol{\psi}_{a1}$ and $\boldsymbol{\psi}_{a2}$, but null values of the energy velocity related to the *optical* waveform $\boldsymbol{\psi}_{a3}$. This result is mathematically consistent with the coincidence between the group and energy velocity. From the quantitative viewpoint, the energy is found to flow with higher velocity through the boundary orthogonal to the propagation direction. Indeed, the velocity $v_r$ is much larger than the velocity $v_t$ for horizontally propagating waves (compare Figures 8a,b), while the opposite occurs for vertically propagating waves (compare Figures 8d,e). As minor remark, the highest velocities of the mechanical energy flow are attained by the second acoustic wave propagating horizontally (maximum $v_r$-value for $\boldsymbol{\psi}_{a2}$ in Figure 8a), which are featured by a large compression factor (recall $\Lambda_P \simeq 1$ for $\boldsymbol{\psi}_{a2}$ in Figure 6d). Again, further parametric analyses allow to verify how the energy velocity orthogonal to the propagation direction ($v_t$ per horizontal waves and $v_r$ for vertical waves) tends to vanish for evanescent material chirality.

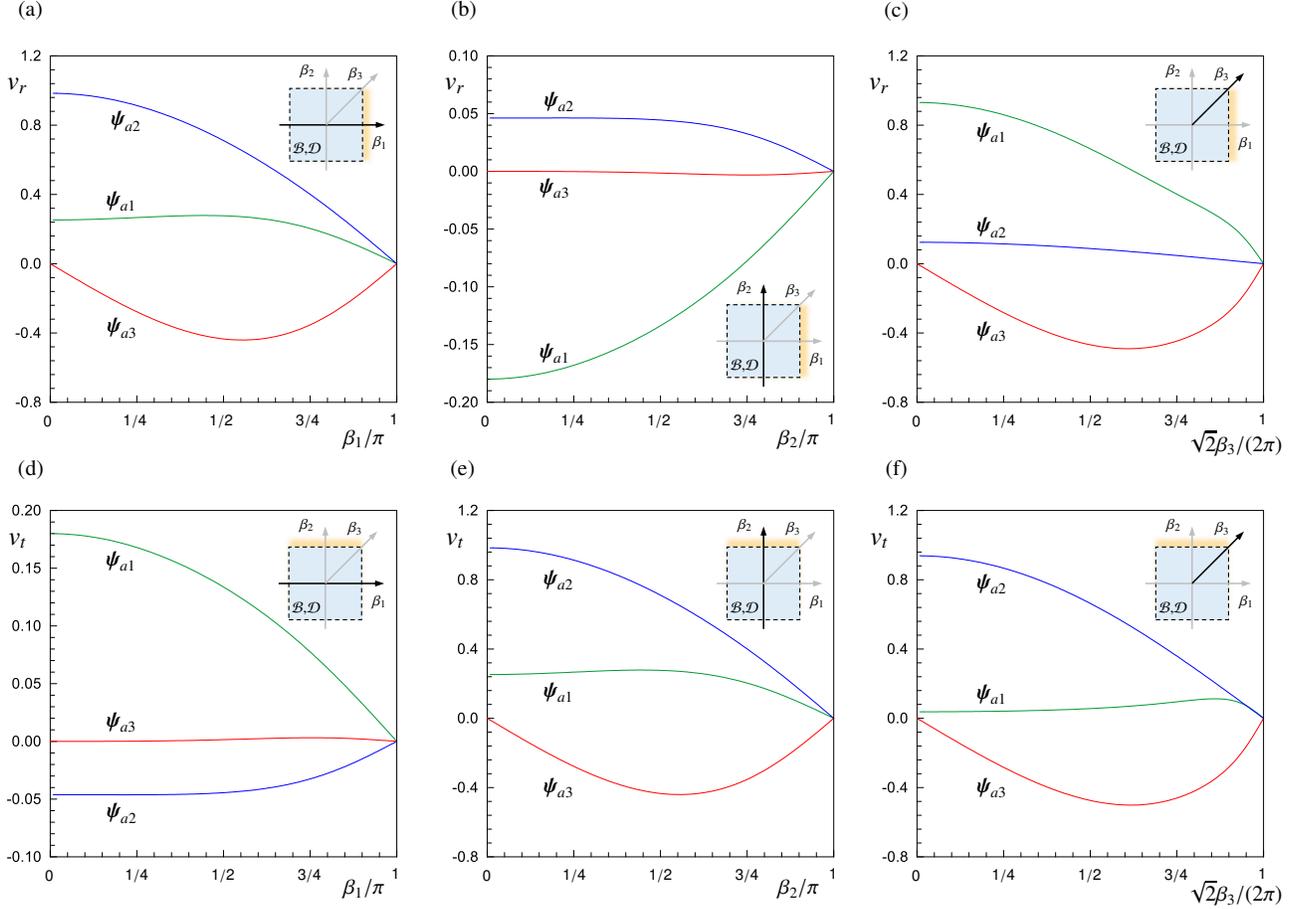

Figure 8: Energy velocity vector $\mathbf{v}_e = (v_r, v_t)$ for the waveforms $\boldsymbol{\psi}_{ai}$ ($i = 1, 2, 3$) of the tetrachiral material with parameter set $\mathbf{p}_*$ ($\delta = \sin\beta = 1/10, \varrho = 1/10, \chi = 1/9$). Velocity of the energy flux flowing out of the right segment $\Gamma_r$ of the cell boundary: (a) $v_r$ versus $\beta_1$ (horizontally propagating waves), (b) $v_r$ versus $\beta_2$ (vertically propagating waves), (c) $v_r$ versus $\beta_3$ (diagonally propagating waves). Velocity of the energy flux flowing out of the top segment $\Gamma_t$ of the cell boundary: (d) $v_t$ versus $\beta_1$ (horizontally propagating waves), (e) $v_t$ versus $\beta_2$ (vertically propagating waves), (f) $v_t$ versus $\beta_3$ (diagonally propagating waves).

The iso-frequency curves related to the dispersion surfaces over the entire first Brillouin zone $\mathcal{B}$ are reported for two tetrachiral materials characterized by *low chirality* (Figure 9a) and *high chirality* (Figure 9b). The vectors related to the phase velocity field $\mathbf{c}_p$ and the group velocity field $\mathbf{c}_g$ are superimposed to the contour plots of the three dispersion surfaces $\omega_1$ (blue), $\omega_2$ (green) and $\omega_3$ (red). Considering a generic $\mathcal{B}$-point, it can be immediately remarked that the velocity vectors $\mathbf{c}_p$ and $\mathbf{c}_g$ are not collinear with each other. On the one hand, the phase velocity vectors $\mathbf{c}_p$ (gray arrows) have amplitude $c_p$ and local direction collinear with the unit wavevector $\tilde{\mathbf{b}}$. This ordinary result means that the wavefront of plane waves propagates along the wavevector direction. On the other hand, the group velocity vectors $\mathbf{c}_g$ are locally orthogonal to the iso-frequency curves. Due to the coincidence between the group and the energy velocities, the $\mathbf{c}_g$-vector field and the $\mathbf{v}_e$-vector field coincide with each other. The physical meaning of this result is that the propagation directions of the envelope wave coincide with the transport directions of the mechanical energy. In synthesis, the mechanical energy flows along directions that may strongly differ from the propagation directions of the dispersive waves. This peculiar behaviour can be attributed to the chiral microstructure of the periodic cell, that corresponds to a marked macroscopic anisotropy of the tetrachiral material. Indeed, the tetrachiral material with higher chirality exhibits lower collinearity between the velocity vectors (compare Figure 9a,b). On the opposite, if the macroscopic anisotropy is reduced by letting the chirality angle tend to zero ($\delta = \sin\beta \to 0$), a cubic material symmetry is established. Therefore, all the dispersion surfaces gain a double symmetry with respect to the coordinate axes of the Brillouin zone $\mathcal{B}$. Along these axes the collinearity of the velocity vectors $\mathbf{c}_p$ and $\mathbf{c}_g$ is respected and the energy flows along the same directions of the wave propagation. Nonetheless, since the cubic material symmetry cannot tend to the isotropy by modifying the other free parameters ($\rho, \chi$), the lack of collinearity persists along all the other directions.

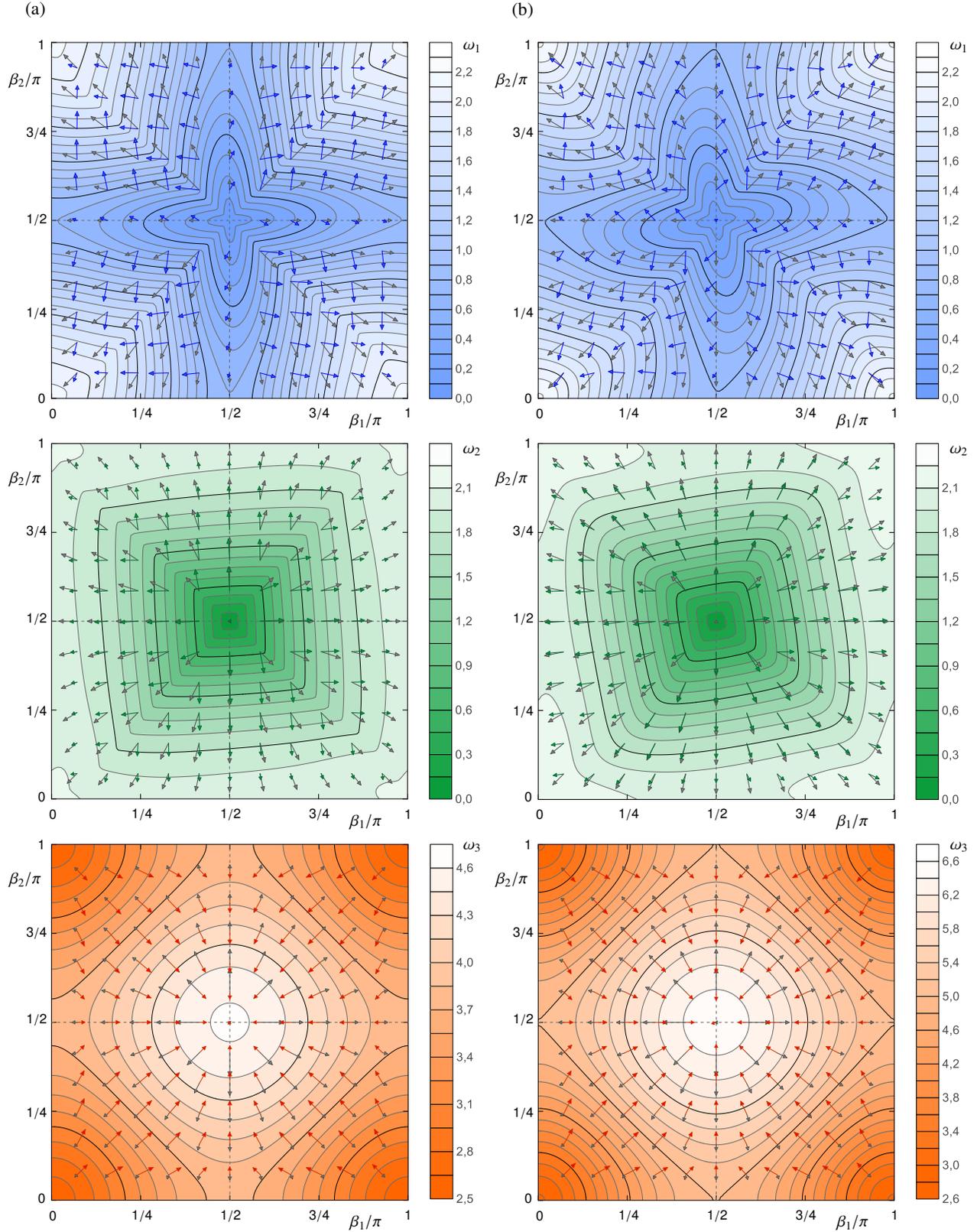

Figure 9: Iso-frequency curves of the dispersion surfaces $\omega_1(\mathbf{b})$ (bluescale), $\omega_2(\mathbf{b})$ (greenscale), $\omega_3(\mathbf{b})$ (redscale) in the Brillouin zone $\mathcal{B}$ for the tetrachiral material (with $\varrho = 1/10, \chi = 1/9$) featured by (a) light chirality (and lower mass density $\delta = \sin\beta = 1/10$) or (b) strong chirality (and higher mass density $\delta = \sin\beta = 1/3$). The gray vectors represent the field of the phase velocity $\mathbf{c}_f$. The blue, green and red vectors represent the field of the group velocity $\mathbf{c}_g$ (and energy velocity $\mathbf{v}_e$).

Focusing on the optical dispersion surface $\omega_3$ (red), it is worth noting that the velocity vectors $\mathbf{c}_p$ and $\mathbf{c}_g$ are oppositely directed ($\mathbf{c}_g \cdot \mathbf{c}_p < 0$) over the whole $\mathcal{B}$-domain. This property is known as *negative refraction* and can equivalently be expressed through the Umov-Poynting vector ($\boldsymbol{\jmath} \cdot \tilde{\mathbf{b}} < 0$). From the physical viewpoint, the occurrence of negative refraction is related to the monotonic decrease of the dispersion surface $\omega_3$ with the distance $\|\mathbf{b}\|$ from the limit point of long wavelengths. This particular acoustic effect is coherent with similar findings from literature studies about periodic lattice structures featured by different tetrachiral microstructures [38].

Focusing on the acoustic dispersion surfaces $\omega_1$ (blue) and $\omega_2$ (green), it can be observed that the $\mathbf{b}$-oriented component of the group velocity tends to the amplitude of the phase velocity ($\mathbf{c}_g \cdot \tilde{\mathbf{b}} \to c_p$) for the limit case of long wavelengths ($\|\mathbf{b}\| \to 0$). This limit behaviour is typical of first-order continua and is consistent with the possibility to approximate the acoustic dispersion surfaces of the beam lattice material with those of a first-order homogenized continuum. For shorter wavelengths the dispersion effects are increasingly important and better approximations of the material spectrum are achievable by homogenization in non-local continua [51, 66]. The boundary of the Brillouin zone $\mathcal{B}$, defining the limit of short wavelengths, is identically featured by vanishing group velocity, but finite not null phase velocities.

## 4. Conclusions

A general mathematical framework has been outlined to analyze the transport of mechanical energy related to the propagation of dispersive acoustic waves in periodic microstructured materials. A non-dissipative beam lattice model, featured by massive rigid bodies connected by massless flexible beams, has been postulated to describe properly the generic microstructure of the periodic cell. A suited dimension reduction has been achieved by applying a classic quasi-static condensation to the *passive* (inertialess) degrees-of-freedom of the external nodes, located at the boundary between adjacent cells. Consequently, standard quasi-periodicity conditions have been imposed to the condensed degrees-of-freedom, according to the Floquet-Bloch theory for conservative systems. Finally, a linear eigenproblem has been stated to govern the free propagation of harmonic plane waves in the reduced space of the *active* degrees-of-freedom.

Since the mechanical energy transfered by freely propagating waves is determined by the complete eigensolution (eigenvalues and eigenvectors), the initial focus has been on the active waveforms related to the dispersion relations of the condensed model. In the specific context of beam lattices, this matter is strictly concerned with the different energetic contents of shear, compression and moment waves. As first original contribution, a pair of nondimensional quantities (*polarization factors*) have been defined to quantify the degree of geometric polarization of a certain wave. Rigorously, perfect polarization may occur in one or the other of the two (mutually-orthogonal) planes extending transversally to the propagation direction. A physically meaningful interpretation is that the *shear* (or *moment*) polarization factor accounts for the ratio between the energy transported by the wavecomponents oscillating (or rotating) transversally to the propagation direction and the total mechanical energy of the waveform. Since each polarization factor can vary continuously from zero to unity, perfectly polarized waves (unitary factor), quasi-polarized waves (nearly-unitary factor) and hybrid waves (two comparable factors) can be distinguished on a consistent quantitative base. A complementary factor has been defined to identify by difference the perfect, quasi-perfect or hybrid *compression* waves, for the sake of completeness.

Drawing inspiration from the Maxwell theory of electromagnetic waves, a vector variable has been introduced to describe the directional flow of mechanical energy associated to the elastic waves propagating in the beam lattices, in analogy to the Umov-Poynting vector in solid mechanics. As major achievement, a mathematically consistent analogy has been established between a local variable of the continuum (the Umov-Poynting vector) and its microstructural counterpart of the beam lattice. Specifically, the periodic cell has been selected as minimal domain for the density of mechanical energy. Therefore, the condensation relations have been employed to express the mechanical energy flowing out the cellular boundary as an explicit function of the active waveforms. As further theoretical development, the directional flow of mechanical energy has been related to the velocity field of the energy transport. Coherently with other literature studies about periodic structures, the energy velocity field has been concluded to coincide with the group velocity field. This major result has been demonstrated by properly inverting the condensation relations to achieve an alternative expression for the group velocity.

Finally, the periodic tetrachiral material has been analyzed as prototypical case-study of beam lattice. The waveforms and the corresponding energy spectrum have been determined exactly. For each waveform, the polarization factors have succeeded in quantitatively following the smooth transition from perfect polarization to quasi-polarization and (more rarely) hybridization, under variation of the wavenumbers. Therefore, the energy fluxes and energy velocities have been determined. As first remark, the highest-energy branches of the dispersion spectrum have been found not immediately associable to the largest energy fluxes, depending on the nonlinear dependence of the energy fluxes on the waveforms. As second remark, the energy velocity (or the group velocity) has been found to differ from the phase velocity in the general case. This difference has been verified to enable the occurrence of negative refractions associated to the optical branch of the dispersion spectrum.

## Acknowledgments

The authors acknowledge financial support of the (MURST) Italian Department for University and Scientific and Technological Research in the framework of the research MIUR Prin15 project 2015LYYXA8, "Multi-scale mechanical models for the design and optimization of micro-structured smart materials and metamaterials", coordinated by prof. A. Corigliano.

# Appendix A. Auxiliary matrices

## Appendix A.1. Periodicity conditions

The $N_p^+ \times N_p^-$ (square) matrix $\mathbf{L(b)}$, governing the periodicity conditions imposed on the passive displacements and forces, can be expressed in the partitioned form

$$\mathbf{L(b)} = \begin{bmatrix} \mathbf{L}_1 & \mathbf{O} \\ \mathbf{O} & \mathbf{L}_2 \end{bmatrix} = \begin{bmatrix} e^{-\iota\beta_1}\mathbf{I}_1 & \mathbf{O} \\ \mathbf{O} & e^{-\iota\beta_2}\mathbf{I}_2 \end{bmatrix} \quad (A.1)$$

where the nondimensional wavevector $\mathbf{b} = (\beta_1, \beta_2)$, while $\mathbf{I}_1$ and $\mathbf{I}_2$ are $N_l \times N_l$ and $N_b \times N_b$ identity matrices, respectively.

## Appendix A.2. Quasi static condensation

According to the boundary segmentation and the consequent partition $\mathbf{q}_p = (\mathbf{q}_p^-, \mathbf{q}_p^+)$ and $\mathbf{f}_p = (\mathbf{f}_p^-, \mathbf{f}_p^+)$ of the passive variables, the quasi-static (lower) part of the equation (2) can be written

$$\begin{bmatrix} \mathbf{K}_{pa}^- \\ \mathbf{K}_{pa}^+ \end{bmatrix} \mathbf{q}_a + \begin{bmatrix} \mathbf{K}_{pp}^= & \mathbf{K}_{pp}^\mp \\ \mathbf{K}_{pp}^\pm & \mathbf{K}_{pp}^\# \end{bmatrix} \begin{pmatrix} \mathbf{q}_p^- \\ \mathbf{q}_p^+ \end{pmatrix} = \begin{pmatrix} \mathbf{f}_p^- \\ \mathbf{f}_p^+ \end{pmatrix} \quad (A.2)$$

where it is worth noting that the submatrices $\mathbf{K}_{pp}^\mp$ e $\mathbf{K}_{pp}^\pm$ express the stiffness coupling between the passive degrees-of-freedom $\mathbf{q}_p^-$ and $\mathbf{q}_p^+$. These stiffness terms vanish in the absence of elastic coupling between the nodes of the boundaries $\Gamma^-$ and $\Gamma^+$.

In combination with the periodicity conditions $\mathbf{q}_p^+ = \mathbf{L}\mathbf{q}_p^-$ and $\mathbf{f}_p^+ = -\mathbf{L}\mathbf{f}_p^-$, the equation (A.2) furnishes the linear laws $\mathbf{q}_p^- = \mathbf{S}_{pa}^-\mathbf{q}_a$ and $\mathbf{f}_p^- = \mathbf{F}_{pp}^-\mathbf{q}_p^- = \mathbf{F}_{pa}^-\mathbf{q}_a$, relating the passive variables $\mathbf{q}_p^-$ and $\mathbf{f}_p^-$ to the active displacements $\mathbf{q}_a$. The rectangular $N_p^- \times N_a$ matrix $\mathbf{S}_{pa}^-$ and the square $N_p^- \times N_p^-$ matrix $\mathbf{F}_{pp}^-$ read

$$\mathbf{S}_{pa}^- = \mathbf{R}(\mathbf{K}_{pa}^+ + \mathbf{L}\mathbf{K}_{pa}^-), \quad (A.3)$$

$$\mathbf{F}_{pp}^- = \mathbf{K}_{pa}^- + (\mathbf{K}_{pp}^= + \mathbf{K}_{pp}^\mp \mathbf{L}) \quad (A.4)$$

and thus, by simple substitution, the $N_p^- \times N_a$ matrix $\mathbf{F}_{pa}^-$ reads

$$\mathbf{F}_{pa}^- = \left(\mathbf{K}_{pa}^- + (\mathbf{K}_{pp}^= + \mathbf{K}_{pp}^\mp \mathbf{L})\right)\mathbf{R}(\mathbf{K}_{pa}^+ + \mathbf{L}\mathbf{K}_{pa}^-) \quad (A.5)$$

where the square matrix $\mathbf{R} = -(\mathbf{L}\mathbf{K}_{pp}^\mp \mathbf{L} + \mathbf{L}\mathbf{K}_{pp}^= + \mathbf{K}_{pp}^\# \mathbf{L} + \mathbf{K}_{pp}^\pm)^{-1}$.

According to the partition of the displacement vector $\mathbf{q}_p^- = (\mathbf{q}_l^-, \mathbf{q}_b^-)$, the matrices $\mathbf{S}_{pa}^-$ and $\mathbf{F}_{pa}^-$ can be partitioned

$$\mathbf{S}_{pa}^- = \begin{bmatrix} \mathbf{S}_{la}^- \\ \mathbf{S}_{ba}^- \end{bmatrix}, \qquad \mathbf{F}_{pa}^- = \begin{bmatrix} \mathbf{F}_{la}^- \\ \mathbf{F}_{ba}^- \end{bmatrix} \quad (A.6)$$

where $\mathbf{S}_{la}^-$ and $\mathbf{S}_{ba}^-$ (and also $\mathbf{F}_{la}^-$ and $\mathbf{F}_{ba}^-$) are $N_l \times N_a$ and $N_b \times N_a$ rectangular matrices, respectively.

Finally, the solution $\mathbf{q}_p^- = \mathbf{S}_{pa}^-\mathbf{q}_a$ and the periodicity condition $\mathbf{q}_p^+ = \mathbf{L}\mathbf{q}_p^-$ can be employed to replace the variable $\mathbf{q}_p$ in the upper part of the equation (2). The algebra leads to the equation (4), governed by the condensed stiffness matrix

$$\mathbf{K}_a = \mathbf{K}_{aa} + (\mathbf{K}_{ap}^- + \mathbf{K}_{ap}^+ \mathbf{L})\mathbf{R}(\mathbf{K}_{pa}^+ + \mathbf{L}\mathbf{K}_{pa}^-) \quad (A.7)$$

which can be proved to be hermitian with dimensions $N_a \times N_a$.

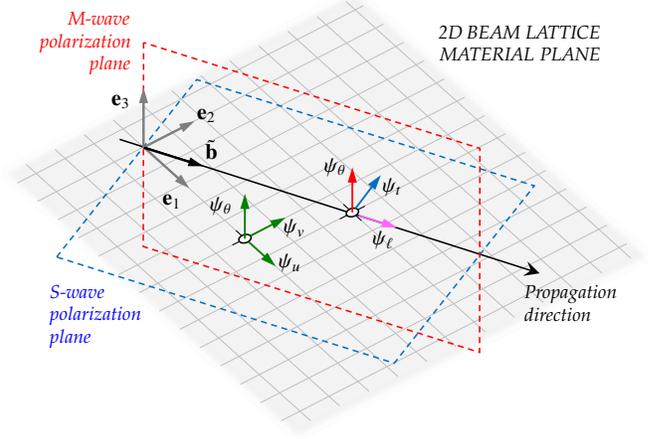

Figure A.10: Graphical illustration of the change of coordinates from the natural basis of the material lattice to the local basis of the plane wave.

## Appendix A.3. Change-of-basis

The skew-symmetric matrix $\mathbf{B}$ employed in the polarization factors (6),(7) accounts for a change-of-basis. It is a diagonal block matrix with $i$-th block ($i = 1, ..., N_n$)

$$\mathbf{B}_i = \begin{bmatrix} \mathbf{e}_1 \cdot \tilde{\mathbf{b}} & \|\mathbf{e}_1 \times \tilde{\mathbf{b}}\| & 0 \\ -\|\mathbf{e}_1 \times \tilde{\mathbf{b}}\| & \mathbf{e}_1 \cdot \tilde{\mathbf{b}} & 0 \\ 0 & 0 & 1 \end{bmatrix} \quad (A.8)$$

which essentially governs the vector transformation $\mathbf{B}_i \tilde{\mathbf{b}} = \mathbf{e}_1$, as illustrated in Figure A.10.

# Appendix B. Umov-Poynting vector

## Appendix B.1. Umov-Poynting Theorem

Considering only mechanical energy, the energy balance for a continuum medium states that the variation rate of the mechanical energy $\mathcal{E}$ is equal to the external mechanical power $\mathcal{P}_E$

$$\frac{\partial \mathcal{E}}{\partial t} = \mathcal{P}_E \quad (B.1)$$

where, considering a generic reference volume $\mathcal{V}$ bounded by the closed surface $\mathcal{S}$, the external mechanical power $\mathcal{P}_E$ due to the body forces $\mathbf{b}$ and surface tractions $\mathbf{f}$ is

$$\mathcal{P}_E = \int_{\mathcal{V}} \mathbf{b} \cdot \dot{\mathbf{u}}\, dV + \int_{\mathcal{S}} \mathbf{f} \cdot \dot{\mathbf{u}}\, dS \quad (B.2)$$

while the increment rate of the mechanical energy is

$$\frac{\partial \mathcal{E}}{\partial t} = \frac{\partial \mathcal{K}}{\partial t} + \mathcal{P}_I = \frac{\partial}{\partial t}\int_{\mathcal{V}} \tfrac{1}{2}\rho \dot{\mathbf{u}} \cdot \dot{\mathbf{u}}\, dV + \int_{\mathcal{V}} \mathbf{T} : \dot{\mathbf{E}}\, dV \quad (B.3)$$

where $\mathcal{P}_I$ denotes the internal mechanical power and $\mathbf{T}$ and $\mathbf{E}$ are the stress and strain tensors. Recalling that the strain rate tensor can be expressed as $\dot{\mathbf{E}} = \nabla\dot{\mathbf{u}} - \text{skw}(\nabla\dot{\mathbf{u}})$ and that $\mathbf{T} : \text{skw}(\nabla \dot{\mathbf{u}}) = 0$ for the symmetry of $\mathbf{T}$, the increment rate of the mechanical energy can be expressed

$$\frac{\partial \mathcal{E}}{\partial t} = \int_{\mathcal{V}} \rho \ddot{\mathbf{u}} \cdot \dot{\mathbf{u}} \, dV + \int_{\mathcal{V}} \mathbf{T} : \nabla \dot{\mathbf{u}} \, dV = \quad (B.4)$$

$$= \int_{\mathcal{V}} \rho \ddot{\mathbf{u}} \cdot \dot{\mathbf{u}} \, dV + \int_{\mathcal{V}} \left( \nabla \cdot (\mathbf{T} \dot{\mathbf{u}}) - (\nabla \cdot \mathbf{T}) \dot{\mathbf{u}} \right) dV = \quad (B.5)$$

$$= \int_{\mathcal{V}} (\rho \ddot{\mathbf{u}} - \nabla \cdot \mathbf{T}) \cdot \dot{\mathbf{u}} \, dV + \int_{\mathcal{S}} (\mathbf{T} \dot{\mathbf{u}}) \cdot \mathbf{n} \, dS \quad (B.6)$$

where $\mathbf{n}$ is the unit outward $\mathcal{S}$-normal. The equation (B.6) is equivalent to the equation (13). In the absence of external body forces ($\mathbf{b} = \mathbf{0}$), the energy balance requires that the $\mathcal{V}$-integral kernel identically vanishes (for all $\mathcal{V}$), which implies $\rho \ddot{\mathbf{u}} = \nabla \cdot \mathbf{T}$. Furthermore, the $\mathcal{S}$-integral must satisfy

$$\int_{\mathcal{S}} (\mathbf{T} \dot{\mathbf{u}}) \cdot \mathbf{n} \, dS = \int_{\mathcal{S}} (\mathbf{T} \mathbf{n}) \cdot \dot{\mathbf{u}} \, dS = \int_{\mathcal{S}} \mathbf{f} \cdot \dot{\mathbf{u}} \, dS \quad (B.7)$$

which requires $(\mathbf{T} \dot{\mathbf{u}}) \cdot \mathbf{n} = \mathbf{f} \cdot \dot{\mathbf{u}}$, or equivalently $\mathbf{s} \cdot \mathbf{n} = -\mathbf{f} \cdot \dot{\mathbf{u}}$ if the Umov-Poynting vector $\mathbf{s} = -\mathbf{T} \dot{\mathbf{u}}$ is employed. It is worth noting that, in the *free oscillation* regime ($\mathcal{P}_E = 0$), the $\mathcal{S}$-surface integral is null (for all $\mathcal{S}$) and can be treated as

$$\int_{\mathcal{S}} (\mathbf{T} \dot{\mathbf{u}}) \cdot \mathbf{n} \, dS = - \int_{\mathcal{S}} (\mathbf{s} \cdot \mathbf{n}) \, dS = - \int_{\mathcal{V}} (\nabla \cdot \mathbf{s}) \, dV \quad (B.8)$$

which is null (for all $\mathcal{V}$) if $\mathbf{s}$ is a solenoidal field with $\nabla \cdot \mathbf{s} = 0$ (see also [1]). From the physical viewpoint, this mathematical condition states that the flux of mechanical energy flowing out any closed surface is null.

*AppendixB.2. Alternative formulation*

An alternative formulation of the Umov-Poynting vector can be given, based on a proper mathematical treatment of the real and imaginary parts of the two harmonic vector fields contributing to the mechanical energy flux defined in equation (21). Separating the real and imaginary parts, the passive forces and the velocities of the passive degrees-of-freedom related to the positive segments of the microstructural boundary $\Gamma^+$ read

$$\dot{\mathbf{q}}_p^+ = \mathfrak{R}(\dot{\mathbf{q}}_p^+) + \iota \mathfrak{I}(\dot{\mathbf{q}}_p^+), \qquad \mathbf{f}_p^+ = \mathfrak{R}(\mathbf{f}_p^+) + \iota \mathfrak{I}(\mathbf{f}_p^+) \quad (B.9)$$

where, employing the relations $\mathbf{q}_p^+ = \mathbf{L} \mathbf{S}_{pa}^- \mathbf{q}_a$ and $\mathbf{f}_p^+ = -\mathbf{L} \mathbf{F}_{pa}^- \mathbf{q}_a$ following from the quasi-periodicity conditions with the quasi-static condensation, the real and imaginary parts read

$$\mathfrak{R}(\dot{\mathbf{q}}_p^+) = \begin{pmatrix} \mathfrak{R}(\mathbf{L}_1 \mathbf{S}_{la}^-) \mathfrak{R}(\dot{\mathbf{q}}_a) - \mathfrak{I}(\mathbf{L}_1 \mathbf{S}_{la}^-) \mathfrak{I}(\dot{\mathbf{q}}_a) \\ \mathfrak{R}(\mathbf{L}_2 \mathbf{S}_{ba}^-) \mathfrak{R}(\dot{\mathbf{q}}_a) - \mathfrak{I}(\mathbf{L}_2 \mathbf{S}_{ba}^-) \mathfrak{I}(\dot{\mathbf{q}}_a) \end{pmatrix} \quad (B.10)$$

$$\mathfrak{I}(\dot{\mathbf{q}}_p^+) = \begin{pmatrix} \mathfrak{R}(\mathbf{L}_1 \mathbf{S}_{la}^-) \mathfrak{I}(\dot{\mathbf{q}}_a) + \mathfrak{I}(\mathbf{L}_1 \mathbf{S}_{la}^-) \mathfrak{R}(\dot{\mathbf{q}}_a) \\ \mathfrak{R}(\mathbf{L}_2 \mathbf{S}_{ba}^-) \mathfrak{I}(\dot{\mathbf{q}}_a) + \mathfrak{I}(\mathbf{L}_2 \mathbf{S}_{ba}^-) \mathfrak{R}(\dot{\mathbf{q}}_a) \end{pmatrix}$$

$$\mathfrak{R}(\mathbf{f}_p^+) = \begin{pmatrix} -\mathfrak{R}(\mathbf{L}_1 \mathbf{F}_{la}^-) \mathfrak{R}(\mathbf{q}_a) + \mathfrak{I}(\mathbf{L}_1 \mathbf{F}_{la}^-) \mathfrak{I}(\mathbf{q}_a) \\ -\mathfrak{R}(\mathbf{L}_2 \mathbf{F}_{ba}^-) \mathfrak{R}(\mathbf{q}_a) + \mathfrak{I}(\mathbf{L}_2 \mathbf{F}_{ba}^-) \mathfrak{I}(\mathbf{q}_a) \end{pmatrix} \quad (B.11)$$

$$\mathfrak{I}(\mathbf{f}_p^+) = \begin{pmatrix} -\mathfrak{R}(\mathbf{L}_1 \mathbf{F}_{la}^-) \mathfrak{I}(\mathbf{q}_a) - \mathfrak{I}(\mathbf{L}_1 \mathbf{F}_{la}^-) \mathfrak{R}(\mathbf{q}_a) \\ -\mathfrak{R}(\mathbf{L}_2 \mathbf{F}_{ba}^-) \mathfrak{I}(\mathbf{q}_a) - \mathfrak{I}(\mathbf{L}_2 \mathbf{F}_{ba}^-) \mathfrak{R}(\mathbf{q}_a) \end{pmatrix}$$

After substitution in equation (21) and expansion of the matrix products, the real-valued components of the Umov-Poynting vector read

$$\mathfrak{R}(J_r(\tau)) = \mathfrak{I}(\mathbf{q}_a)^\top \mathfrak{I}(\mathbf{L}_1 \mathbf{F}_{la}^-)^\top \mathfrak{I}(\mathbf{L}_1 \mathbf{S}_{la}^-) \mathfrak{I}(\dot{\mathbf{q}}_a) + \quad (B.12)$$
$$- \mathfrak{I}(\mathbf{q}_a)^\top \mathfrak{I}(\mathbf{L}_1 \mathbf{F}_{la}^-)^\top \mathfrak{R}(\mathbf{L}_1 \mathbf{S}_{la}^-) \mathfrak{R}(\dot{\mathbf{q}}_a) +$$
$$- \mathfrak{R}(\mathbf{q}_a)^\top \mathfrak{R}(\mathbf{L}_1 \mathbf{F}_{la}^-)^\top \mathfrak{I}(\mathbf{L}_1 \mathbf{S}_{la}^-) \mathfrak{I}(\dot{\mathbf{q}}_a) +$$
$$+ \mathfrak{R}(\mathbf{q}_a)^\top \mathfrak{R}(\mathbf{L}_1 \mathbf{F}_{la}^-)^\top \mathfrak{R}(\mathbf{L}_1 \mathbf{S}_{la}^-) \mathfrak{R}(\dot{\mathbf{q}}_a)$$

$$\mathfrak{R}(J_t(\tau)) = \mathfrak{I}(\mathbf{q}_a)^\top \mathfrak{I}(\mathbf{L}_2 \mathbf{F}_{ba}^-)^\top \mathfrak{I}(\mathbf{L}_2 \mathbf{S}_{ba}^-) \mathfrak{I}(\dot{\mathbf{q}}_a) + \quad (B.13)$$
$$- \mathfrak{I}(\mathbf{q}_a)^\top \mathfrak{I}(\mathbf{L}_2 \mathbf{F}_{ba}^-)^\top \mathfrak{R}(\mathbf{L}_2 \mathbf{S}_{ba}^-) \mathfrak{R}(\dot{\mathbf{q}}_a) +$$
$$- \mathfrak{R}(\mathbf{q}_a)^\top \mathfrak{R}(\mathbf{L}_2 \mathbf{F}_{ba}^-)^\top \mathfrak{I}(\mathbf{L}_2 \mathbf{S}_{ba}^-) \mathfrak{I}(\dot{\mathbf{q}}_a) +$$
$$+ \mathfrak{R}(\mathbf{q}_a)^\top \mathfrak{R}(\mathbf{L}_2 \mathbf{F}_{ba}^-)^\top \mathfrak{R}(\mathbf{L}_2 \mathbf{S}_{ba}^-) \mathfrak{R}(\dot{\mathbf{q}}_a)$$

where the real and imaginary parts of the matrix products can be expressed in the trigonometric form

$$\mathfrak{R}(\mathbf{L}_1 \mathbf{S}_{la}^-) = \mathfrak{R}(\mathbf{S}_{la}^-) \cos \beta_1 + \mathfrak{I}(\mathbf{S}_{la}^-) \sin \beta_1 \quad (B.14)$$
$$\mathfrak{I}(\mathbf{L}_1 \mathbf{S}_{la}^-) = \mathfrak{I}(\mathbf{S}_{la}^-) \cos \beta_1 - \mathfrak{R}(\mathbf{S}_{la}^-) \sin \beta_1$$
$$\mathfrak{R}(\mathbf{L}_2 \mathbf{S}_{ba}^-) = \mathfrak{R}(\mathbf{S}_{ba}^-) \cos \beta_2 + \mathfrak{I}(\mathbf{S}_{ba}^-) \sin \beta_2$$
$$\mathfrak{I}(\mathbf{L}_2 \mathbf{S}_{ba}^-) = \mathfrak{I}(\mathbf{S}_{ba}^-) \cos \beta_2 - \mathfrak{R}(\mathbf{S}_{ba}^-) \sin \beta_2$$
$$\mathfrak{R}(\mathbf{L}_1 \mathbf{F}_{la}^-) = \mathfrak{R}(\mathbf{F}_{la}^-) \cos \beta_1 + \mathfrak{I}(\mathbf{F}_{la}^-) \sin \beta_1$$
$$\mathfrak{I}(\mathbf{L}_1 \mathbf{F}_{la}^-) = \mathfrak{I}(\mathbf{F}_{la}^-) \cos \beta_1 - \mathfrak{R}(\mathbf{F}_{la}^-) \sin \beta_1$$
$$\mathfrak{R}(\mathbf{L}_2 \mathbf{F}_{ba}^-) = \mathfrak{R}(\mathbf{F}_{ba}^-) \cos \beta_2 + \mathfrak{I}(\mathbf{F}_{ba}^-) \sin \beta_2$$
$$\mathfrak{I}(\mathbf{L}_2 \mathbf{F}_{ba}^-) = \mathfrak{I}(\mathbf{F}_{ba}^-) \cos \beta_2 - \mathfrak{R}(\mathbf{F}_{ba}^-) \sin \beta_2$$

while, recalling the natural wave solution $\mathbf{q}_a = A \boldsymbol{\psi}_a \exp(\iota \omega \tau)$, the real and imaginary parts of the displacement and velocity vectors can be expressed in the trigonometric form

$$\mathfrak{R}(\mathbf{q}_a) = \left( \mathfrak{R}(A) \mathfrak{R}(\boldsymbol{\psi}_a) - \mathfrak{I}(A) \mathfrak{I}(\boldsymbol{\psi}_a) \right) \cos(\omega \tau) +$$
$$- \left( \mathfrak{I}(A) \mathfrak{R}(\boldsymbol{\psi}_a) + \mathfrak{R}(A) \mathfrak{I}(\boldsymbol{\psi}_a) \right) \sin(\omega \tau) \quad (B.15)$$

$$\mathfrak{I}(\mathbf{q}_a) = \left( \mathfrak{R}(A) \mathfrak{I}(\boldsymbol{\psi}_a) + \mathfrak{I}(A) \mathfrak{R}(\boldsymbol{\psi}_a) \right) \cos(\omega \tau) +$$
$$- \left( \mathfrak{I}(A) \mathfrak{I}(\boldsymbol{\psi}_a) - \mathfrak{R}(A) \mathfrak{R}(\boldsymbol{\psi}_a) \right) \sin(\omega \tau) \quad (B.16)$$

$$\mathfrak{R}(\dot{\mathbf{q}}_a) = -\omega \left( \mathfrak{R}(A) \mathfrak{I}(\boldsymbol{\psi}_a) + \mathfrak{I}(A) \mathfrak{R}(\boldsymbol{\psi}_a) \right) \cos(\omega \tau) +$$
$$- \omega \left( \mathfrak{R}(A) \mathfrak{R}(\boldsymbol{\psi}_a) - \mathfrak{I}(A) \mathfrak{I}(\boldsymbol{\psi}_a) \right) \sin(\omega \tau) \quad (B.17)$$

$$\mathfrak{I}(\dot{\mathbf{q}}_a) = -\omega \left( \mathfrak{I}(A) \mathfrak{I}(\boldsymbol{\psi}_a) + \mathfrak{R}(A) \mathfrak{R}(\boldsymbol{\psi}_a) \right) \cos(\omega \tau) +$$
$$- \omega \left( \mathfrak{R}(A) \mathfrak{I}(\boldsymbol{\psi}_a) + \mathfrak{I}(A) \mathfrak{R}(\boldsymbol{\psi}_a) \right) \sin(\omega \tau) \quad (B.18)$$

Similarly, the real-valued components of the *mean* flux of mechanical energy defined in equation (23) read

$$\bar{J}_r = \tfrac{AA^*}{2} \omega \left( \mathfrak{R}(\boldsymbol{\psi}_a)^\top \mathbf{Y}_r^a \mathfrak{R}(\boldsymbol{\psi}_a) + \mathfrak{I}(\boldsymbol{\psi}_a)^\top \mathbf{Y}_r^a \mathfrak{I}(\boldsymbol{\psi}_a) + \right. \quad (B.19)$$
$$\left. - \mathfrak{R}(\boldsymbol{\psi}_a)^\top \mathbf{Y}_r^b \mathfrak{I}(\boldsymbol{\psi}_a) + \mathfrak{I}(\boldsymbol{\psi}_a)^\top \mathbf{Y}_r^b \mathfrak{R}(\boldsymbol{\psi}_a) \right)$$

$$\bar{J}_t = \tfrac{AA^*}{2} \omega \left( \mathfrak{R}(\boldsymbol{\psi}_a)^\top \mathbf{Y}_t^a \mathfrak{R}(\boldsymbol{\psi}_a) + \mathfrak{I}(\boldsymbol{\psi}_a)^\top \mathbf{Y}_t^a \mathfrak{I}(\boldsymbol{\psi}_a) + \right. \quad (B.20)$$
$$\left. - \mathfrak{R}(\boldsymbol{\psi}_a)^\top \mathbf{Y}_t^b \mathfrak{I}(\boldsymbol{\psi}_a) + \mathfrak{I}(\boldsymbol{\psi}_a)^\top \mathbf{Y}_t^b \mathfrak{R}(\boldsymbol{\psi}_a) \right)$$

where the auxiliary matrices

$$\begin{aligned}\mathbf{Y}_r^a &= \mathfrak{I}(\mathbf{L}_1\mathbf{F}_{la}^-)^\top \mathfrak{R}(\mathbf{L}_1\mathbf{S}_{la}^-) + \mathfrak{R}(\mathbf{L}_1\mathbf{F}_{la}^-)^\top \mathfrak{I}(\mathbf{L}_1\mathbf{S}_{la}^-) = \\ &= \mathfrak{I}(\mathbf{F}_{la}^-)^\top \mathfrak{R}(\mathbf{S}_{la}^-) - \mathfrak{R}(\mathbf{F}_{la}^-)^\top \mathfrak{I}(\mathbf{S}_{la}^-)\end{aligned} \quad (B.21)$$

$$\begin{aligned}\mathbf{Y}_r^b &= \mathfrak{R}(\mathbf{L}_1\mathbf{F}_{la}^-)^\top \mathfrak{R}(\mathbf{L}_1\mathbf{S}_{la}^-) + \mathfrak{I}(\mathbf{L}_1\mathbf{F}_{la}^-)^\top \mathfrak{I}(\mathbf{L}_1\mathbf{S}_{la}^-) = \\ &= \mathfrak{I}(\mathbf{F}_{la}^-)^\top \mathfrak{I}(\mathbf{S}_{la}^-) + \mathfrak{R}(\mathbf{F}_{la}^-)^\top \mathfrak{R}(\mathbf{S}_{la}^-)\end{aligned} \quad (B.22)$$

$$\begin{aligned}\mathbf{Y}_t^a &= \mathfrak{I}(\mathbf{L}_2\mathbf{F}_{ba}^-)^\top \mathfrak{R}(\mathbf{L}_2\mathbf{S}_{ba}^-) + \mathfrak{R}(\mathbf{L}_2\mathbf{F}_{ba}^-)^\top \mathfrak{I}(\mathbf{L}_2\mathbf{S}_{ba}^-) = \\ &= \mathfrak{I}(\mathbf{F}_{ba}^-)^\top \mathfrak{R}(\mathbf{S}_{ba}^-) - \mathfrak{R}(\mathbf{F}_{ba}^-)^\top \mathfrak{I}(\mathbf{S}_{ba}^-)\end{aligned} \quad (B.23)$$

$$\begin{aligned}\mathbf{Y}_t^b &= \mathfrak{R}(\mathbf{L}_2\mathbf{F}_{ba}^-)^\top \mathfrak{R}(\mathbf{L}_2\mathbf{S}_{ba}^-) + \mathfrak{I}(\mathbf{L}_2\mathbf{F}_{ba}^-)^\top \mathfrak{I}(\mathbf{L}_2\mathbf{S}_{ba}^-) = \\ &= \mathfrak{I}(\mathbf{F}_{ba}^-)^\top \mathfrak{I}(\mathbf{S}_{ba}^-) + \mathfrak{R}(\mathbf{F}_{ba}^-)^\top \mathfrak{R}(\mathbf{S}_{ba}^-)\end{aligned} \quad (B.24)$$

## Appendix C. Energy relations

The mechanical energy $\mathcal{E}$ in the periodic cell of a beam lattice material is the sum $\mathcal{E} = \mathcal{K} + \mathcal{U}$, where $\mathcal{K}$ and $\mathcal{U}$ are the kinetic and (internal) potential energies. Considering a harmonic wave motion $\mathbf{q}(\tau)$ with complex waveform, the mean energy values yielding from proper averaging over the time period are

$$\bar{\mathcal{K}} = \frac{\omega}{2\pi}\int_0^{\frac{2\pi}{\omega}} \tfrac{1}{2}\,\mathfrak{R}(\dot{\mathbf{q}})^\top \mathbf{M}\,\mathfrak{R}(\dot{\mathbf{q}})\, d\tau, \quad (C.1)$$

$$\bar{\mathcal{U}} = \frac{\omega}{2\pi}\int_0^{\frac{2\pi}{\omega}} \tfrac{1}{2}\,\mathfrak{R}(\mathbf{q})^\top \mathbf{K}\,\mathfrak{R}(\mathbf{q})\, d\tau \quad (C.2)$$

which are equivalent to

$$\bar{\mathcal{K}} = \tfrac{1}{2}\mathfrak{R}\left(\dot{\mathbf{q}}^\top \mathbf{M}\,\dot{\mathbf{q}}^*\right), \qquad \bar{\mathcal{U}} = \tfrac{1}{2}\mathfrak{R}\left(\mathbf{q}^\top \mathbf{K}\,\mathbf{q}^*\right) \quad (C.3)$$

where $\dot{\mathbf{q}}$ and $\mathbf{q}$ are the velocity and displacement vectors of all the configuration nodes, while the products $\mathbf{M}\dot{\mathbf{q}}$ and $\mathbf{K}\mathbf{q}$ stand for the momentum the elastic force vectors, respectively.

Recalling the displacement partition $\mathbf{q} = (\mathbf{q}_a, \mathbf{q}_p)$ and the consequent partition of the mass stiffness matrices $\mathbf{M}$ and $\mathbf{K}$ in the equation (2), the kinetic energy can be expressed

$$\bar{\mathcal{K}} = \tfrac{1}{2}\mathfrak{R}\left(\dot{\mathbf{q}}_a^\top \mathbf{M}_a \dot{\mathbf{q}}_a^*\right) = \tfrac{1}{8}\left(\dot{\mathbf{q}}_a^\top \mathbf{M}_a \dot{\mathbf{q}}_a^* + \dot{\mathbf{q}}_a^\dagger \mathbf{M}_a \dot{\mathbf{q}}_a\right) \quad (C.4)$$

where the quadratic dependence on the velocity of the only active degrees-of-freedom can be recognized. Similarly, if the displacement partition is introduced, the potential energy reads

$$\begin{aligned}\bar{\mathcal{U}} = &\tfrac{1}{2}\mathfrak{R}(\mathbf{q}_a^\top \mathbf{K}_{aa}\mathbf{q}_a^*) + \tfrac{1}{2}\mathfrak{R}(\mathbf{q}_a^\top \mathbf{K}_{ap}\mathbf{q}_p^*) + \\ &+ \tfrac{1}{2}\mathfrak{R}(\mathbf{q}_p^\top \mathbf{K}_{pa}\mathbf{q}_a^*) + \tfrac{1}{2}\mathfrak{R}(\mathbf{q}_p^\top \mathbf{K}_{pp}\mathbf{q}_p^*)\end{aligned} \quad (C.5)$$

Therefore, imposing both the quasi-static condensation and the quasi-periodicity conditions on the passive degrees-of-freedom, the potential energy can be expressed

$$\bar{\mathcal{U}} = \tfrac{1}{2}\mathfrak{R}(\mathbf{q}_a^\top \mathbf{K}_a^* \mathbf{q}_a^*) + \tfrac{1}{2}\mathfrak{R}(\mathbf{q}_p^\top \mathbf{K}_{pa}\mathbf{q}_a^*) + \tfrac{1}{2}\mathfrak{R}(\mathbf{q}_p^\top \mathbf{K}_{pp}\mathbf{q}_p^*) \quad (C.6)$$

where two terms can be recognized

$$\bar{\mathcal{U}}_a = \tfrac{1}{2}\mathfrak{R}(\mathbf{q}_a^\top \mathbf{K}_a^* \mathbf{q}_a^*) \quad (C.7)$$

$$\bar{\mathcal{U}}_p = \tfrac{1}{2}\mathfrak{R}(\mathbf{q}_p^\top \mathbf{K}_{pa}\mathbf{q}_a^*) + \tfrac{1}{2}\mathfrak{R}(\mathbf{q}_p^\top \mathbf{K}_{pp}\mathbf{q}_p^*) \quad (C.8)$$

corresponding to the potential energy stored in the active and passive degrees-of-freedom, respectively.

The equilibrium condition (2), after multiplication of the upper and lower parts by $\mathbf{q}_a$ and $\mathbf{q}_p$, respectively, and averaging over a period, states that

$$\bar{\mathcal{U}}_a = \bar{\mathcal{K}}, \qquad \bar{\mathcal{U}}_p = \tfrac{1}{2}\bar{\mathcal{W}} \quad (C.9)$$

where $\bar{\mathcal{W}}$ is the mean work done by the passive external forces $\mathbf{f}_p$ in the passive degrees-of-freedom

$$\bar{\mathcal{W}} = \frac{\omega}{2\pi}\int_0^{\frac{2\pi}{\omega}} \mathfrak{R}(\mathbf{q}_p)^\top \mathfrak{R}(\mathbf{f}_p)\, d\tau \quad (C.10)$$

which is equivalent to

$$\bar{\mathcal{W}} = \mathfrak{R}\left(\mathbf{q}_p^\top \mathbf{f}_p^*\right) = \tfrac{1}{4}\left(\mathbf{q}_p^\top \mathbf{f}_p^* + \mathbf{q}_p^\dagger \mathbf{f}_p\right) \quad (C.11)$$

Therefore, the internal potential energy can be expressed

$$\bar{\mathcal{U}} = \bar{\mathcal{U}}_a + \bar{\mathcal{U}}_p = \bar{\mathcal{K}} + \tfrac{1}{2}\bar{\mathcal{W}} \quad (C.12)$$

and the mean mechanical energy $\bar{\mathcal{E}}$ is

$$\bar{\mathcal{E}} = \bar{\mathcal{K}} + \bar{\mathcal{U}} = (\bar{\mathcal{K}} + \bar{\mathcal{U}}_a) + \bar{\mathcal{U}}_p = 2\bar{\mathcal{K}} + \tfrac{1}{2}\bar{\mathcal{W}} \quad (C.13)$$

where the two contributions given by the active and passive degrees-of-freedom have been separated.

The work done by the passive external forces $\mathbf{f}_p$ in the passive degrees-of-freedom can be separated by distinguishing the contributions $\bar{\mathcal{W}}^-$ and $\bar{\mathcal{W}}^+$ stored in the negative and positive microstructural boundaries $\Gamma^-$ and $\Gamma^+$, respectively

$$\bar{\mathcal{W}} = \bar{\mathcal{W}}^- + \bar{\mathcal{W}}^+ \quad (C.14)$$

where, recalling the displacement partition $\mathbf{q}_p = (\mathbf{q}_p^-, \mathbf{q}_p^+)$, the two distinct contributions read

$$\bar{\mathcal{W}}^- = \tfrac{1}{4}\left((\mathbf{q}_p^-)^\top (\mathbf{f}_p^-)^* + (\mathbf{q}_p^-)^\dagger (\mathbf{f}_p^-)\right) \quad (C.15)$$

$$\bar{\mathcal{W}}^+ = \tfrac{1}{4}\left((\mathbf{q}_p^+)^\top (\mathbf{f}_p^+)^* + (\mathbf{q}_p^+)^\dagger (\mathbf{f}_p^+)\right) \quad (C.16)$$

Therefore, imposing the quasi-periodicity conditions $\mathbf{q}_p^+ = \mathbf{L}\mathbf{q}_p^-$ and $\mathbf{f}_p^+ = -\mathbf{L}\mathbf{f}_p^-$, the work done by the passive forces in the positive boundary $\Gamma^+$ can be expressed

$$\begin{aligned}\bar{\mathcal{W}}^+ &= -\tfrac{1}{4}\left((\mathbf{L}\mathbf{q}_p^-)^\top (\mathbf{L}\mathbf{f}_p^-)^* + (\mathbf{L}\mathbf{q}_p^-)^\dagger (\mathbf{L}\mathbf{f}_p^-)\right) = \\ &= -\tfrac{1}{4}\left((\mathbf{q}_p^-)^\top \mathbf{L}^\top \mathbf{L}^* (\mathbf{f}_p^-)^* + (\mathbf{q}_p^-)^\dagger \mathbf{L}^\dagger \mathbf{L}\mathbf{f}_p^-\right)\end{aligned} \quad (C.17)$$

where the products $\mathbf{L}^\top \mathbf{L}^* = \mathbf{L}^\dagger \mathbf{L} = \mathbf{I}$ by construction. In conclusion, the work done by the passive forces is null

$$\bar{\mathcal{W}} = \bar{\mathcal{W}}^- + \bar{\mathcal{W}}^+ = 0 \quad (C.18)$$

since the work done by the passive forces in the negative boundary $\Gamma^+$ is exactly balanced by that done by the passive forces in the positive boundary. Thus, the equation (43) is verified.

## Appendix D. Group velocity

The alternative equations (50) and (51) for the group velocity can be demonstrated by opportune $\beta_i$-differentiation of the condensed and non-condensed equations of motion, respectively, according to the following mathematical procedures.

## AppendixD.1. Stiffness derivative formulation

The equation (50) can be demonstrated by $\beta_i$-differentiation of the condensed equation of motion (4), yielding

$$\partial_{\beta_i}\left(\left(\mathbf{K}_a - \omega^2 \mathbf{M}_a\right)\boldsymbol{\psi}_a\right) = \mathbf{0} \quad \text{(D.1)}$$

The left-hand term can be developed in the form

$$\left(\partial_{\beta_i}\mathbf{K}_a - \partial_{\beta_i}(\omega^2)\mathbf{M}_a - \omega^2\partial_{\beta_i}\mathbf{M}_a\right)\boldsymbol{\psi}_a + \left(\mathbf{K}_a - \omega^2\mathbf{M}_a\right)\partial_{\beta_i}\boldsymbol{\psi}_a \quad \text{(D.2)}$$

where the mass matrix is $\beta_i$-independent and $\partial_{\beta_i}\mathbf{M}_a = \mathbf{O}$. Therefore, the equation (D.1) becomes

$$\left(\partial_{\beta_i}\mathbf{K}_a - \partial_{\beta_i}(\omega^2)\mathbf{M}_a\right)\boldsymbol{\psi}_a + \left(\mathbf{K}_a - \omega^2\mathbf{M}_a\right)\partial_{\beta_i}\boldsymbol{\psi}_a = \mathbf{0} \quad \text{(D.3)}$$

Pre-multiplication left and right terms by $\boldsymbol{\psi}_a^\dagger$ gives

$$\boldsymbol{\psi}_a^\dagger\left(\partial_{\beta_i}\mathbf{K}_a - \partial_{\beta_i}(\omega^2)\mathbf{M}_a\right)\boldsymbol{\psi}_a + \boldsymbol{\psi}_a^\dagger\left(\mathbf{K}_a - \omega^2\mathbf{M}_a\right)\partial_{\beta_i}\boldsymbol{\psi}_a = 0 \quad \text{(D.4)}$$

whose complex conjugate, after simple algebraic manipulations, assumes the form

$$\boldsymbol{\psi}_a^\dagger\left(\partial_{\beta_i}\mathbf{K}_a - \partial_{\beta_i}(\omega^2)\mathbf{M}_a\right)^\dagger\boldsymbol{\psi}_a + \left(\partial_{\beta_i}\boldsymbol{\psi}_a\right)^\dagger\left(\mathbf{K}_a - \omega^2\mathbf{M}_a\right)^\dagger\boldsymbol{\psi}_a = 0 \quad \text{(D.5)}$$

Recalling that $\omega$ is real-valued, $\mathbf{K}_a$ is a hermitian matrix by construction and $\mathbf{M}_a$ is a real and symmetric matrix, the following identities can be demonstrated

$$\left(\mathbf{K}_a - \omega^2\mathbf{M}_a\right)^\dagger\boldsymbol{\psi}_a = \left(\mathbf{K}_a - \omega^2\mathbf{M}_a\right)\boldsymbol{\psi}_a = \mathbf{0} \quad \text{(D.6)}$$

Consequently, the first term of the equation (D.5) gives

$$\boldsymbol{\psi}_a^\dagger\left(\partial_{\beta_i}\mathbf{K}_a\right)^\dagger\boldsymbol{\psi}_a = \partial_{\beta_i}(\omega^2)\boldsymbol{\psi}_a^\dagger\mathbf{M}_a\boldsymbol{\psi}_a \quad \text{(D.7)}$$

and finally

$$\partial_{\beta_i}(\omega^2) = 2\omega\,\partial_{\beta_i}\omega = \frac{\boldsymbol{\psi}_a^\dagger\partial_{\beta_i}\mathbf{K}_a\boldsymbol{\psi}_a}{\boldsymbol{\psi}_a^\dagger\mathbf{M}_a\boldsymbol{\psi}_a} \quad \text{(D.8)}$$

where the following identity has been employed

$$\left(\partial_{\beta_i}\mathbf{K}_a\right)^\dagger = \partial_{\beta_i}\mathbf{K}_a \quad \text{(D.9)}$$

which derives from hermitian property of $\mathbf{K}_a$. The equation (50) follows immediately from the relation (D.8). Since the numerator and denominator of the right-hand term in the relation (D.8) are real-valued due to the properties of complex conjugate eigenvectors, their ratio is real-valued, as required.

## AppendixD.2. Waveform derivative formulation

The equation (51) can be demonstrated by $\beta_i$-differentiation of the non-condensed equation of motion (1), where the harmonic monofrequent solution $\mathbf{q} = \boldsymbol{\psi}\exp(\iota\omega\tau)$ and $\mathbf{f} = \boldsymbol{\varphi}\exp(\iota\omega\tau)$ are imposed, where $\boldsymbol{\psi} = (\boldsymbol{\psi}_a, \boldsymbol{\psi}_p)$ e $\boldsymbol{\varphi} = (\mathbf{0}, \boldsymbol{\varphi}_p)$. After substitution and removing the $\tau$-dependence, the equation reads

$$\mathbf{K}\boldsymbol{\psi} - \omega^2\mathbf{M}\boldsymbol{\psi} = \boldsymbol{\varphi} \quad \text{(D.10)}$$

Pre-multiplication left and right terms by $\boldsymbol{\psi}^\dagger$ gives

$$\boldsymbol{\psi}^\dagger\mathbf{K}\boldsymbol{\psi} - \omega^2\boldsymbol{\psi}^\dagger\mathbf{M}\boldsymbol{\psi} = \boldsymbol{\psi}^\dagger\boldsymbol{\varphi} \quad \text{(D.11)}$$

Applying $\beta_i$-differentiation to both terms, the following equation is obtained

$$\partial_{\beta_i}\left(\boldsymbol{\psi}^\dagger\mathbf{K}\boldsymbol{\psi}\right) - \partial_{\beta_i}\left(\omega^2\boldsymbol{\psi}^\dagger\mathbf{M}\boldsymbol{\psi}\right) = \partial_{\beta_i}\left(\boldsymbol{\psi}^\dagger\boldsymbol{\varphi}\right) \quad \text{(D.12)}$$

which can be manipulated according to the differentiation properties to achieve

$$\partial_{\beta_i}\left(\boldsymbol{\psi}^\dagger\right)\mathbf{K}\boldsymbol{\psi} + \boldsymbol{\psi}^\dagger\partial_{\beta_i}\mathbf{K}\boldsymbol{\psi} + \boldsymbol{\psi}^\dagger\mathbf{K}\partial_{\beta_i}\boldsymbol{\psi} - \partial_{\beta_i}\left(\omega^2\right)\boldsymbol{\psi}^\dagger\mathbf{M}\boldsymbol{\psi} +$$
$$-\omega^2\partial_{\beta_i}\left(\boldsymbol{\psi}^\dagger\right)\mathbf{M}\boldsymbol{\psi} - \omega^2\boldsymbol{\psi}^\dagger\partial_{\beta_i}\mathbf{M}\boldsymbol{\psi} - \omega^2\boldsymbol{\psi}^\dagger\mathbf{M}\partial_{\beta_i}\boldsymbol{\psi} =$$
$$= \partial_{\beta_i}\left(\boldsymbol{\psi}^\dagger\right)\boldsymbol{\varphi} + \boldsymbol{\psi}^\dagger\partial_{\beta_i}\boldsymbol{\varphi} \quad \text{(D.13)}$$

Recalling that the matrices $\mathbf{K}$ and $\mathbf{M}$ are $\beta_i$-independent, the equation (D.13) becomes

$$\partial_{\beta_i}\left(\boldsymbol{\psi}^\dagger\right)\mathbf{K}\boldsymbol{\psi} + \boldsymbol{\psi}^\dagger\mathbf{K}\partial_{\beta_i}\boldsymbol{\psi} - \partial_{\beta_i}\left(\omega^2\right)\boldsymbol{\psi}^\dagger\mathbf{M}\boldsymbol{\psi} +$$
$$-\omega^2\partial_{\beta_i}\left(\boldsymbol{\psi}^\dagger\right)\mathbf{M}\boldsymbol{\psi} - \omega^2\boldsymbol{\psi}^\dagger\mathbf{M}\partial_{\beta_i}\boldsymbol{\psi} =$$
$$= \partial_{\beta_i}\left(\boldsymbol{\psi}^\dagger\right)\boldsymbol{\varphi} + \boldsymbol{\psi}^\dagger\partial_{\beta_i}\boldsymbol{\varphi} \quad \text{(D.14)}$$

or, in a more significant form

$$\partial_{\beta_i}\left(\boldsymbol{\psi}^\dagger\right)\left(\mathbf{K}\boldsymbol{\psi} - \omega^2\mathbf{M}\boldsymbol{\psi}\right) + \boldsymbol{\psi}^\dagger\left(\mathbf{K}\partial_{\beta_i}\boldsymbol{\psi} - \omega^2\mathbf{M}\partial_{\beta_i}\boldsymbol{\psi}\right) +$$
$$-\partial_{\beta_i}\left(\omega^2\right)\boldsymbol{\psi}^\dagger\mathbf{M}\boldsymbol{\psi} = \partial_{\beta_i}\left(\boldsymbol{\psi}^\dagger\right)\boldsymbol{\varphi} + \boldsymbol{\psi}^\dagger\partial_{\beta_i}\boldsymbol{\varphi} \quad \text{(D.15)}$$

Simple algebraic manipulation allow to modify the equation in the convenient form

$$\partial_{\beta_i}\left(\boldsymbol{\psi}^\dagger\right)\left(\mathbf{K}\boldsymbol{\psi} - \omega^2\mathbf{M}\boldsymbol{\psi}\right) + \left(\partial_{\beta_i}\left(\boldsymbol{\psi}^\dagger\right)\left(\mathbf{K}\boldsymbol{\psi} - \omega^2\mathbf{M}\boldsymbol{\psi}\right)\right)^\dagger +$$
$$-\partial_{\beta_i}\left(\omega^2\right)\boldsymbol{\psi}^\dagger\mathbf{M}\boldsymbol{\psi} = \partial_{\beta_i}\left(\boldsymbol{\psi}^\dagger\right)\boldsymbol{\varphi} + \boldsymbol{\psi}^\dagger\partial_{\beta_i}\boldsymbol{\varphi} \quad \text{(D.16)}$$

where the equation (D.10) can be recalled to state that

$$\partial_{\beta_i}\left(\boldsymbol{\psi}^\dagger\right)\boldsymbol{\varphi} + \left(\partial_{\beta_i}\left(\boldsymbol{\psi}^\dagger\right)\boldsymbol{\varphi}\right)^\dagger +$$
$$-\partial_{\beta_i}\left(\omega^2\right)\boldsymbol{\psi}^\dagger\mathbf{M}\boldsymbol{\psi} = \partial_{\beta_i}\left(\boldsymbol{\psi}^\dagger\right)\boldsymbol{\varphi} + \boldsymbol{\psi}^\dagger\partial_{\beta_i}\boldsymbol{\varphi} \quad \text{(D.17)}$$

which can be algebrically manipulated to obtain the relation

$$\partial_{\beta_i}\left(\omega^2\right) = 2\omega\,\partial_{\beta_i}\omega = \frac{\boldsymbol{\varphi}_p^\dagger\partial_{\beta_i}\boldsymbol{\psi}_p - \boldsymbol{\psi}_p^\dagger\partial_{\beta_i}\boldsymbol{\varphi}_p}{\boldsymbol{\psi}_a^\dagger\mathbf{M}_a\boldsymbol{\psi}_a} \quad \text{(D.18)}$$

where the partitions $\boldsymbol{\psi} = (\boldsymbol{\psi}_a, \boldsymbol{\psi}_p)$ and $\boldsymbol{\varphi} = (\mathbf{0}, \boldsymbol{\varphi}_p)$, together with the corresponding matrix partitions, have been employed. The equation (51) follows immediately from the relation (D.18).

## AppendixD.3. Group velocity versus energy velocity

The two terms contributing to the numerator of equation (51) for the group velocity can be handled according to algebraic and differential manipulations

$$\begin{aligned}\boldsymbol{\varphi}_p^\dagger \partial_{\beta_i}\boldsymbol{\psi}_p &= (\boldsymbol{\varphi}_p^-)^\dagger \partial_{\beta_i}\boldsymbol{\psi}_p^- + (\boldsymbol{\varphi}_p^+)^\dagger (\partial_{\beta_i}\boldsymbol{\psi}_p^+) = \\ &= (\boldsymbol{\varphi}_p^-)^\dagger \partial_{\beta_i}\boldsymbol{\psi}_p^- + (-\mathbf{L}\boldsymbol{\varphi}_p^-)^\dagger \partial_{\beta_i}(\mathbf{L}\boldsymbol{\psi}_p^-) = \\ &= (\boldsymbol{\varphi}_p^-)^\dagger \partial_{\beta_i}\boldsymbol{\psi}_p^- + (-\mathbf{L}\boldsymbol{\varphi}_p^-)^\dagger \partial_{\beta_i}\mathbf{L}\boldsymbol{\psi}_p^- + (-\mathbf{L}\boldsymbol{\varphi}_p^-)^\dagger \mathbf{L}\partial_{\beta_i}\boldsymbol{\psi}_p^- = \\ &= (\boldsymbol{\varphi}_p^-)^\dagger \partial_{\beta_i}\boldsymbol{\psi}_p^- - (\boldsymbol{\varphi}_p^-)^\dagger \mathbf{L}^\dagger \partial_{\beta_i}\mathbf{L}\boldsymbol{\psi}_p^- - (\boldsymbol{\varphi}_p^-)^\dagger \mathbf{L}^\dagger \mathbf{L}\partial_{\beta_i}\boldsymbol{\psi}_p^- = \\ &= (\boldsymbol{\varphi}_p^-)^\dagger (\mathbf{I}-\mathbf{L}^\dagger\mathbf{L})\partial_{\beta_i}\boldsymbol{\psi}_p^- - (\boldsymbol{\varphi}_p^-)^\dagger \mathbf{L}^\dagger \partial_{\beta_i}\mathbf{L}\boldsymbol{\psi}_p^- = \\ &= -(\boldsymbol{\varphi}_p^-)^\dagger \mathbf{L}^\dagger \partial_{\beta_i}\mathbf{L}\boldsymbol{\psi}_p^- \end{aligned} \quad (D.19)$$

$$\begin{aligned}\boldsymbol{\psi}_p^\dagger \partial_{\beta_i}\boldsymbol{\varphi}_p &= (\boldsymbol{\psi}_p^-)^\dagger \partial_{\beta_i}\boldsymbol{\varphi}_p^- + (\boldsymbol{\psi}_p^+)^\dagger \partial_{\beta_i}\boldsymbol{\varphi}_p^+ = \\ &= (\boldsymbol{\psi}_p^-)^\dagger \partial_{\beta_i}\boldsymbol{\varphi}_p^- - (\mathbf{L}\boldsymbol{\psi}_p^-)^\dagger \partial_{\beta_i}(\mathbf{L}\boldsymbol{\varphi}_p^-) = \\ &= (\boldsymbol{\psi}_p^-)^\dagger \partial_{\beta_i}\boldsymbol{\varphi}_p^- - (\mathbf{L}\boldsymbol{\psi}_p^-)^\dagger \partial_{\beta_i}\mathbf{L}\boldsymbol{\varphi}_p^- - (\mathbf{L}\boldsymbol{\psi}_p^-)^\dagger \mathbf{L}\partial_{\beta_i}\boldsymbol{\varphi}_p^- = \\ &= (\boldsymbol{\psi}_p^-)^\dagger \partial_{\beta_i}\boldsymbol{\varphi}_p^- - (\boldsymbol{\psi}_p^-)^\dagger \mathbf{L}^\dagger \partial_{\beta_i}\mathbf{L}\boldsymbol{\varphi}_p^- - (\boldsymbol{\psi}_p^-)^\dagger \mathbf{L}^\dagger \mathbf{L}\partial_{\beta_i}\boldsymbol{\varphi}_p^- = \\ &= (\boldsymbol{\psi}_p^-)^\dagger (\mathbf{I}-\mathbf{L}^\dagger\mathbf{L})\partial_{\beta_i}\boldsymbol{\varphi}_p^- - (\boldsymbol{\psi}_p^-)^\dagger \mathbf{L}^\dagger \partial_{\beta_i}\mathbf{L}\boldsymbol{\varphi}_p^- = \\ &= -(\boldsymbol{\psi}_p^-)^\dagger \mathbf{L}^\dagger \partial_{\beta_i}\mathbf{L}\boldsymbol{\varphi}_p^- \end{aligned} \quad (D.20)$$

Therefore, distinguishing between the partial derivatives with respect to one or the other wavenumbers, the terms of the equations (D.19),(D.20) can be specified

$$\boldsymbol{\varphi}_p^\dagger \partial_{\beta_1}\boldsymbol{\psi}_p = -(\boldsymbol{\varphi}_p^-)^\dagger \mathbf{L}^\dagger \partial_{\beta_1}\mathbf{L}\boldsymbol{\psi}_p^- = \iota(\boldsymbol{\varphi}_l^-)^\dagger \boldsymbol{\psi}_l^- \quad (D.21)$$

$$\boldsymbol{\varphi}_p^\dagger \partial_{\beta_2}\boldsymbol{\psi}_p = -(\boldsymbol{\varphi}_p^-)^\dagger \mathbf{L}^\dagger \partial_{\beta_2}\mathbf{L}\boldsymbol{\psi}_p^- = \iota(\boldsymbol{\varphi}_b^-)^\dagger \boldsymbol{\psi}_b^- \quad (D.22)$$

$$\boldsymbol{\psi}_p^\dagger \partial_{\beta_1}\boldsymbol{\varphi}_p = -(\boldsymbol{\psi}_p^-)^\dagger \mathbf{L}^\dagger \partial_{\beta_1}\mathbf{L}\boldsymbol{\varphi}_p^- = \iota(\boldsymbol{\psi}_l^-)^\dagger \boldsymbol{\varphi}_l^- \quad (D.23)$$

$$\boldsymbol{\psi}_p^\dagger \partial_{\beta_2}\boldsymbol{\varphi}_p = -(\boldsymbol{\psi}_p^-)^\dagger \mathbf{L}^\dagger \partial_{\beta_2}\mathbf{L}\boldsymbol{\varphi}_p^- = \iota(\boldsymbol{\psi}_b^-)^\dagger \boldsymbol{\varphi}_b^- \quad (D.24)$$

The substitution of the relations (D.21)-(D.24) into the equations (D.19),(D.20) gives the following expressions for the group velocity components

$$c_{g1} = \frac{\iota}{2\omega} \frac{(\boldsymbol{\varphi}_l^-)^\dagger \boldsymbol{\psi}_l^- - (\boldsymbol{\psi}_l^-)^\dagger \boldsymbol{\varphi}_l^-}{\boldsymbol{\psi}_a^\dagger \mathbf{M}_a \boldsymbol{\psi}_a} \quad (D.25)$$

$$c_{g2} = \frac{\iota}{2\omega} \frac{(\boldsymbol{\varphi}_b^-)^\dagger \boldsymbol{\psi}_b^- - (\boldsymbol{\psi}_b^-)^\dagger \boldsymbol{\varphi}_b^-}{\boldsymbol{\psi}_a^\dagger \mathbf{M}_a \boldsymbol{\psi}_a} \quad (D.26)$$

and, employing the properties $(\boldsymbol{\psi}_l^-)^\dagger \boldsymbol{\varphi}_l^- = ((\boldsymbol{\psi}_l^-)^\dagger \boldsymbol{\varphi}_l^-)^\top$ and $(\boldsymbol{\psi}_b^-)^\dagger \boldsymbol{\varphi}_b^- = ((\boldsymbol{\psi}_b^-)^\dagger \boldsymbol{\varphi}_b^-)^\top$ at the numerator, the group velocities becomes

$$c_{g1} = \frac{\iota}{2\omega} \frac{(\boldsymbol{\varphi}_l^-)^\dagger \boldsymbol{\psi}_l^- - (\boldsymbol{\varphi}_l^-)^\top (\boldsymbol{\psi}_l^-)^*}{\boldsymbol{\psi}_a^\dagger \mathbf{M}_a \boldsymbol{\psi}_a} \quad (D.27)$$

$$c_{g2} = \frac{\iota}{2\omega} \frac{(\boldsymbol{\varphi}_b^-)^\dagger \boldsymbol{\psi}_b^- - (\boldsymbol{\varphi}_b^-)^\top (\boldsymbol{\psi}_b^-)^*}{\boldsymbol{\psi}_a^\dagger \mathbf{M}_a \boldsymbol{\psi}_a} \quad (D.28)$$

Recalling the important relations $\boldsymbol{\psi}_l^- = \mathbf{S}_{la}^- \boldsymbol{\psi}_a$, $\boldsymbol{\varphi}_l^- = \mathbf{F}_{la}^- \boldsymbol{\psi}_a$, $\boldsymbol{\psi}_b^- = \mathbf{S}_{ba}^- \boldsymbol{\psi}_a$, $\boldsymbol{\varphi}_b^- = \mathbf{F}_{ba}^- \boldsymbol{\psi}_a$, the group velocity components read

$$c_{g1} = \frac{\iota}{2\omega} \frac{(\mathbf{F}_{la}^- \boldsymbol{\psi}_a)^\dagger \mathbf{S}_{la}^- \boldsymbol{\psi}_a - (\mathbf{F}_{la}^- \boldsymbol{\psi}_a)^\top (\mathbf{S}_{la}^- \boldsymbol{\psi}_a)^*}{\boldsymbol{\psi}_a^\dagger \mathbf{M}_a \boldsymbol{\psi}_a} \quad (D.29)$$

$$c_{g2} = \frac{\iota}{2\omega} \frac{(\mathbf{F}_{ba}^- \boldsymbol{\psi}_a)^\dagger \mathbf{S}_{ba}^- \boldsymbol{\psi}_a - (\mathbf{F}_{ba}^- \boldsymbol{\psi}_a)^\top (\mathbf{S}_{ba}^- \boldsymbol{\psi}_a)^*}{\boldsymbol{\psi}_a^\dagger \mathbf{M}_a \boldsymbol{\psi}_a} \quad (D.30)$$

which are equivalent to the energy velocity components (46),(47).

# AppendixE. Physical matrices of the tetrachiral material

## AppendixE.1. Non condensed model

The non-null coefficients $M_{ij}$ of the mass submatrix $\mathbf{M}_a$ (with dimensions $3\times 3$) and the non-null coefficients $K_{ij}$ of the stiffness matrix $\mathbf{K}$ (with dimensions $15\times 15$) governing the free dynamics of the tetrachiral material read

- submatrix $\mathbf{M}_a$ with dimensions $3\times 3$ and $i,j=1...3$

$$M_{11} = M_{22} = 1, \quad M_{33} = \chi^2 \quad (E.1)$$

- submatrix $\mathbf{K}_{aa}$ with dimensions $3\times 3$ and $i,j=1...3$

$$K_{11} = K_{22} = 4d_5(48\varrho^2 + \Delta^2) \quad (E.2)$$
$$K_{33} = 2d_3(16\varrho^2 + \Delta^2\delta^2)$$

- submatrix $\mathbf{K}_{ap}$ with dimensions $3\times 12$ and $i=1...3$ and $j=4...15$

$$K_{14} = K_{17} = -2d_5(48\varrho^2\delta^2 + \Delta^3) \quad (E.3)$$
$$K_{15} = K_{18} = K_{111} = K_{114} = 2d_4\delta(48\varrho^2 - \Delta^2)$$
$$K_{16} = -K_{19} = -24d_4\delta\varrho^2$$
$$K_{110} = K_{113} = -2d_3(48\varrho^2 + \Delta\delta^2)$$
$$K_{112} = -K_{115} = 24d_3\varrho^2$$
$$K_{24} = K_{27} = -K_{210} = -K_{213} = -2d_4\delta(48\varrho^2 - \Delta^2)$$
$$K_{25} = K_{28} = -2d_3(48\varrho^2 + \Delta\delta^2)$$
$$K_{26} = -K_{29} = -24d_3\varrho^2$$
$$K_{211} = K_{214} = -2d_5(48\varrho^2\delta^2 + \Delta^3)$$
$$K_{212} = -K_{215} = -24d_4\delta\varrho^2$$
$$K_{34} = -K_{37} = K_{311} = -K_{314} = d_4\delta(24\varrho^2 - \Delta^2)$$
$$K_{35} = -K_{38} = -K_{310} = K_{313} = d_3(24\varrho^2 + \Delta\delta^2)$$
$$K_{36} = K_{39} = K_{312} = K_{315} = 4d_3\varrho^2$$

- submatrix $\mathbf{K}_{pp}$ with dimensions $12\times 12$ and $i=4...15$ and $j=4...15$

$$K_{44} = K_{77} = K_{1111} = K_{1414} = 2d_5(48\varrho^2\delta^2 + \Delta^3) \quad (E.4)$$
$$K_{55} = K_{88} = K_{1010} = K_{1313} = 2d_3(48\varrho^2 + \Delta\delta^2)$$
$$K_{66} = K_{99} = K_{1212} = K_{1515} = 8d_3\varrho^2$$
$$K_{45} = K_{78} = -K_{1011} = -K_{1314} = 2d_4\delta(48\varrho^2 - \Delta^2)$$
$$K_{46} = -K_{79} = K_{1112} = -K_{1415} = 24d_4\delta\varrho^2$$
$$K_{56} = -K_{89} = -K_{1012} = K_{1315} = 24d_3\varrho^2$$

where $\mathbf{K}_{ap} = \mathbf{K}_{pa}^\top$ and $\mathbf{K}_{pp} = \mathbf{K}_{pp}^\top$ for the symmetric property of the stiffness matrix $\mathbf{K}$.

*Appendix E.2. Condensed model*

The non-null coefficients $K_{aij}$ ($i,j = 1, 2, 3$) of the 3-by-3 condensed stiffness matrix $\mathbf{K}_a(\mathbf{p}, \mathbf{b})$ governing the wave equation for the tetrachiral material read

$$\begin{aligned}
K_{a11} &= c_1 - c_2 \cos\beta_1 - c_3 \cos\beta_2 \\
K_{a22} &= c_1 - c_3 \cos\beta_1 - c_2 \cos\beta_2 \\
K_{a33} &= c_4 + c_5(\cos\beta_1 + \cos\beta_2) \\
K_{a12} &= 2c_6(\cos\beta_1 - \cos\beta_2) \\
K_{a13} &= -\iota\,(c_6 \sin\beta_1 + c_7 \sin\beta_2) \\
K_{a23} &= \iota\,(c_7 \sin\beta_1 - c_6 \sin\beta_2)
\end{aligned} \qquad \text{(E.5)}$$

where the mass coefficients must be divided by $\omega_c^2$ if the normalization frequency $\omega_c$ is not unitary and

$$\begin{aligned}
c_1 &= 2d_5\left(\Delta^2 + 12\varrho^2\right), & c_2 &= 2d_5\left(\Delta^3 + 12\delta^2\varrho^2\right) \\
c_3 &= 2d_3\left(\Delta\delta^2 + 12\varrho^2\right), & c_4 &= d_3\left(\Delta\delta^2 + 16\varrho^2\right) \\
c_5 &= \tfrac{1}{2}d_3\left(\Delta\delta^2 + 8\varrho^2\right), & c_6 &= d_4\delta\left(\Delta^2 - 12\varrho^2\right) \\
c_7 &= d_3\left(\Delta\delta^2 + 12\varrho^2\right)
\end{aligned} \qquad \text{(E.6)}$$

are fully geometric parameters. The auxiliary parameters

$$\begin{aligned}
\Delta &= (1-\delta^2), & d_4 &= 1/(1-\delta^2)^2 \\
d_3 &= 1/(1-\delta^2)^{3/2}, & d_5 &= 1/(1-\delta^2)^{5/2}
\end{aligned} \qquad \text{(E.7)}$$

have been introduced for the sake of conciseness. Finally, the coefficients $K_{aji}$ are the complex conjugate of $K_{aij}$ by virtue of the Hermitian property of the matrix $\mathbf{K}_a$.